\documentclass[12pt]{article}
\usepackage{graphicx}
\usepackage{amsmath}
\usepackage{amssymb}
\usepackage{subfigure}
\usepackage{latexsym}
\usepackage{hyperref}
\usepackage[sort]{cite}
\usepackage{bbm}

\newcommand{\be}{\begin{eqnarray}}
\newcommand{\ee}{\end{eqnarray}}
\newcommand{\bc}{\begin{center}}
\newcommand{\ec}{\end{center}}
\newcommand{\bea}{\begin{eqnarray}}
\newcommand{\eea}{\end{eqnarray}}
\newcommand{\ben}{\begin{equation}}

\newcommand{\nn}{\nonumber}

\numberwithin{equation}{section}

\newsavebox{\ns}
\newsavebox{\dbrane}
\newsavebox{\dbshort}

\usepackage{epsfig}
\usepackage{amssymb}

\def\appendix{{\newpage\section*{Appendix}}\let\appendix\section%
        {\setcounter{section}{0}
        \gdef\thesection{\Alph{section}}}\section}

\newcommand\ba{\begin{eqnarray}}
\newcommand\ea{\end{eqnarray}}

\def\Dslash{\,\,{\raise.15ex\hbox{/}\mkern-12mu D}}
\def\Dbarslash{\,\,{\raise.15ex\hbox{/}\mkern-12mu {\bar D}}}
\def\delslash{\,\,{\raise.15ex\hbox{/}\mkern-9mu \partial}}
\def\delbarslash{\,\,{\raise.15ex\hbox{/}\mkern-9mu {\bar\partial}}}
\def\pslash{\,\,{\raise.15ex\hbox{/}\mkern-9mu p}}
\def\calDslash{\,\,{\raise.15ex\hbox{/}\mkern-12mu {\cal D}}}

\newcommand{\hh}{{1\over 2}}

\renewcommand{\ll}{_}
\newcommand{\uu}{^}
\newcommand{\pp}{\partial}
\renewcommand{\L}{\Lambda}
\renewcommand{\exp}[1]{{\rm exp}\left( #1 \right)}
\newcommand{\expr}[1]{{\rm exp}\left ( #1 \right ) }
\renewcommand{\d}{\delta}
\newcommand{\m}{\mu}

\renewcommand{\m}{\mu}
\newcommand{\n}{\nu}
\newcommand{\s}{\sigma}

\newcommand{\G}{\Gamma}
\newcommand{\g}{\gamma}
\renewcommand{\a}{\alpha}

\renewcommand{\o}{\omega}
\newcommand{\e}{\epsilon}

\newcommand{\sqd}{^2}

\renewcommand{\hh}{{1\over 2}}
\renewcommand{\gg}{\nabla}

\newcommand{\eee}[1]{\ba{#1}\ea}
\renewcommand{\th}{\theta}

\renewcommand{\b}{\beta}

\newcommand{\pr}{^\prime {}}

\newcommand{\apr}{{\alpha^\prime} {}}

\newcommand{\IZ}{\relax\ifmmode\mathchoice
{\hbox{\cmss Z\kern-.4em Z}}{\hbox{\cmss Z\kern-.4em Z}}
{\lower.9pt\hbox{\cmsss Z\kern-.4em Z}} {\lower1.2pt\hbox{\cmsss
Z\kern-.4em Z}}\else{\cmss Z\kern-.4em Z}\fi} \font\cmss=cmss10
\font\cmsss=cmss10 at 7pt
\newcommand{\inbar}{\,\vrule height1.5ex width.4pt depth0pt}
\newcommand{\IC}{{\relax\hbox{$\inbar\kern-.3em{\rm C}$}}}
\newcommand{\IQ}{{\relax\hbox{$\inbar\kern-.3em{\rm Q}$}}}
\newcommand{\IP}{\relax{\rm I\kern-.18em P}}
\newcommand{\Ione}{{\relax\hbox{$\inbar\kern-.39em{\rm 1}$}}}

\newcommand{\ed}{\dot{e}}

\renewcommand{\l}{\lambda}

\newcommand{\cc}{{\cal C}}

\renewcommand{\cc}{{c_1}}

\renewcommand{\o}{\omega}

\newcommand{\ct}{\tilde{c}}

\renewcommand{\cc}{c}

\renewcommand{\L}{\Lambda}
\renewcommand{\pr}{{}^\prime{}}

\newcommand{\pst}{\tilde{\psi}}

\newcommand{\cl}{{\cal L}}

\newcommand{\IR}{\relax{\rm I\kern-.18em R}}
\def\blfootnote{\xdef\@thefnmark{}\@footnotetext}

\renewcommand{\cc}[1]{\cite{#1}}

\newcommand{\bm}{\begin{matrix}}

\newcommand{\rr}[1]{(\ref{{#1}})}
\newcommand{\bbb}{\ba}
\renewcommand{\eee}{\ea}
\newcommand{\een}[1]{\label{#1}\ea}

\newcommand{\xxx}{(\xx)}

\newcommand{\Xd}{{\dot{X}}}

\newcommand{\Xdd}{{\ddot{X}}}

\def\lrdd{\left( \,  }
\def\rrdd{\, \right )}
\def\lsqq{\left [ \, }
\def\rsqq{\, \right ]}

\def\bi{\begin{itemize}}
\def\ei{\end{itemize}}

\def\ed{\end{document}}

\def\cc{{\cal C}}

\renewcommand{\rr}[1]{(\ref{#1})}

\def\ct{{\cal T}}

\def\numosc{{N}}
\def\fin{^{\rm (fin.)}}
\def\rws{R_{\rm ws}}
\def\llt{\tilde{\l}}
\def\cc{\,}
\def\mflw{(-1)\uu{\rm F_{\rm L_{\rm w}}}}
\def\mfw{(-1)\uu{\rm F_{\rm w}}}

\def\xxx{\nn\eee\bbb}

\def\vwsol{V_{\rm ws}^{\rm 1-loop}}
\def\vwscl{V_{\rm ws}^{\rm classical}}
\def\vwsrem{V_{\rm ws}^{\rm remainder}}
\begin{document}

\begin{titlepage}
\begin{flushright}
hep-th/0612051
\end{flushright}
\vspace{15 mm}
\begin{center}
{\Large \bf Dimension-changing exact solutions of string theory}  
\end{center}
\vspace{6 mm}
\begin{center}
{ Simeon Hellerman and Ian Swanson }\\
\vspace{6mm}
{\it School of Natural Sciences, Institute for Advanced Study\\
Princeton, NJ 08540, USA }
\end{center}
\vspace{6 mm}
\begin{center}
{\large Abstract}
\end{center}
\noindent
Superstring theories in the critical dimension $D=10$ are connected 
to one another by a well-explored web of dualities. In this paper we 
use closed-string tachyon condensation 
to connect the supersymmetric moduli space of the critical superstring
to non-supersymmetric string theories in more than ten dimensions. 
We present a new set of classical solutions
that exhibit dynamical transitions between string theories in different
dimensions, with different degrees of stability and different amounts of spacetime
supersymmetry.  In all examples, the string-frame metric and
dilaton gradient readjust themselves during the transition.
The central charge of the worldsheet theory remains equal to $15$, 
even as the total number of dimensions changes.  
This phenomenon arises entirely from a one-loop diagram on the string worldsheet.
Allowed supersymmetric final states include half-BPS vacua of type II 
and $SO(32)$ heterotic string theory.
We also find solutions that bypass the critical dimension altogether 
and proceed directly to spacelike linear dilaton theories in dimensions 
greater than or equal to two.
\vspace{1cm}
\begin{flushleft}
December 7, 2006
\end{flushleft}
\end{titlepage}
\tableofcontents
\newpage

\section{Introduction}
Weakly coupled string theories in $D>10$ dimensions are never supersymmetric,
and they are described by an effective action with a positive 
potential energy.  It is not possible to have a
supersymmetric state of a theory in which the vacuum energy is positive.
In $D \geq 12$ general considerations
would appear to forbid unbroken supersymmetry: any maximally symmetric space 
with $12$ or more dimensions, Lorentzian signature, 
and unbroken supersymmetry must have at least $64$ 
fermionic generators in the unbroken superalgebra.  Acting on massless states such
as the graviton, the superalgebra would force the inclusion of massless states with
helicity $\geq 5/2$.  No consistent interacting effective theory with massless
states of helicity greater than $2$ is known, and such theories are believed
not to exist (at least not with a Lorentz-invariant vacuum).\footnote{For early
work on the inconsistency of massless higher-spin fields, see for example \cite{Aragone:1979hx}.}

In $D=11$, the unique consistent low-energy effective theory with
a supersymmetric ground state is 11-dimensional supergravity, which has
a 32-component Majorana spinor's worth of fermionic generators.  This
theory appears as the effective action of a point on
the supersymmetric moduli space of string theory; it becomes 
infinitely strongly coupled when considered as a string theory.

Nonetheless, weakly coupled string theories in dimensions $11,12,\cdots $ can be
formulated, and appear to be consistent internally.  Many such
theories have tachyonic modes whose condensation can reduce the number of spacetime
dimensions dynamically \cite{hellerman2}.  It would be interesting to know whether
this process can connect a theory in $D>10$ to
the web of supersymmetric vacua in $10$ dimensions.  Similarly, it would be useful
to study whether one can reduce the number of dimensions dynamically in a
critical or subcritical superstring theory to reach a ground state with a
smaller number of dimensions.
It has been argued that this may be the case \cite{hellerman1}.  
If one
treats the worldsheet theory semiclassically, it is easy to find two-dimensional theories
describing the propagation of
a string in a background where the number of spatial dimensions in the far future
is different from that in the far past.  
The resulting worldsheet theories are
super-renormalizable theories wherein
dimensionful couplings are dressed with exponentials of the time coordinate to
render them scale-invariant. 

Such two-dimensional 
field theories define solutions of string theory only if they are 
conformally invariant at the quantum level.  It has been checked \cite{hellerman2} that
conformal invariance holds quantum mechanically to leading nontrivial order in ${1/ D}$,
in the limit where the number of spacetime dimensions becomes large and the 
change $\Delta D$ in the number of spacetime dimensions is held fixed.  In this limit,
worldsheet quantum corrections to classical quantities are suppressed by the quantity 
$D\uu{-(\#~{\rm loops})}$.  This result was subsequently generalized in \cite{freedman}.

While the large-$D$ limit provides a check on the consistency of dynamical 
dimension change in
supercritical string theory, we are still left to wonder whether it is possible
to reach the critical dimension by closed string tachyon condensation.  When the number of
dimensions approaches $10$, the expansion parameter ${1/ D}$ is no longer parametrically
small, and quantum corrections to the semiclassical worldsheet picture are no longer suppressed.
In the absence of general existence theorems for such a CFT, the only hope is to
\it find exact solutions \rm describing closed string tachyon condensation of the kind
that reduces the number of spatial dimensions dynamically.

In this paper we describe a set of such exact solutions.  As in
\cite{previous}, the models in this paper describe tachyon
condensation along a null direction $X\uu +$, rather than a timelike direction.  
In the model of \cite{previous}, the
closed string tachyon was independent of the $D-2$ dimensions
transverse to $X\uu\pm$.  (The exact conformal invariance
of those backgrounds was also remarked upon in \cite{atnulltach1, atnulltach2}.)
In the models described
in the present paper, the
tachyon has a linear or quadratic
dependence on a third coordinate $X\ll 2$,
as well as exponential dependence on $X\uu +$.
The effect is that string states are either pushed out along the
phase boundary at $X\uu + \simeq 0$, or else confined to the
point in the $X\ll 2$ dimension where the tachyon reaches its minimum.
At late times, strings inside the tachyon condensate are confined
to $X\ll 2 = 0$ with a
linear restoring force that increases exponentially with
time.  As a result, strings
are no longer free to move in the $X\ll 2$ direction,
and the number of spacetime dimensions decreases
effectively by one. This model can be generalized
to the case where the number of spacetime dimensions
decreases by any amount $\Delta D$.

The physics is qualitatively
similar to that of \cite{hellerman2},
but quantum corrections are exactly calculable
without taking $D\to\infty$.  In particular,
we can study the
case where the number of dimensions
in the far future can be equal to
the critical dimension or less.
When the final dimension is equal to the
critical dimension, the dilaton gradient is
lightlike, rolling toward weak coupling
in the future.

The plan of the paper is as follows.  
In Section 2, we describe a 
dimension-changing bubble solution\footnote{In \cite{previous},
it was argued that the lightlike Liouville wall can be thought of as the
late-time limit of an expanding bubble wall, as illustrated in Fig.~6 below.} 
in the bosonic string and use 
it to illustrate some aspects of dimension-changing
bubbles in general.  In Section 3, we consider dimension-changing bubbles
in unstable heterotic string theories, with special attention given to the
case in which the final state preserves spacetime supersymmetry in the asymptotic future,
verifying along the way a conjecture made in \cite{hellerman1} and explored further in
\cite{hellerman2}.
In Section 4 we introduce the corresponding solutions in the type 0 string,
including solutions in which the asymptotic future is a supersymmetric
vacuum of type II string theory.  In Section 5 we discuss related issues and
conclusions.

\section{Dimension-changing bubble in the bosonic string}
We focus in this section on classical solutions of bosonic string theory that 
describe transitions between theories of different spacetime
dimension.  Although we defer a more detailed description to a
separate paper, we will rely on a few crucial elements of these solutions
to characterize several aspects of the dynamics.

\subsection{Giving the bosonic string tachyon $\ct$ an expectation value}
In \cite{previous}, we described a background of the timelike linear
dilaton theory of the bosonic string in which the tachyon $\ct$ condensed
along a lightlike direction:
\bbb
\ct(X) = \m\sqd \cc\exp{\b X\uu +}\ .
\eee
A constant dilaton gradient $\pp\ll\m \Phi \equiv V\ll\m$
must satisfy
\bbb
V\ll\m V\uu\m = - {{D - 26}\over{6\apr}}\ ,
\eee
and the on-shell condition for a tachyon perturbation takes the form
\bbb
\pp\uu\m \pp\ll\m \ct - 2 V\uu\m \pp\ll \m \ct + {4\over\apr} \ct = 0\ .
\eee
For a timelike linear dilaton $\Phi = - q X\uu 0$,
the exponent $\b$ is
determined by $\b q = {{2\sqrt{2}}\over\apr}$, and the magnitude $q$
is $q\sqd = {1\over{6\apr}} (D - 26)$.  This background can be
thought of as the late-time limit of an expanding bubble of a
$\ct > 0$ phase of the tachyon.  Nothing can propagate
deeply into the $\ct > 0$ phase, not even gravitons.  
The $\ct > 0$ phase can be understood to represent an
absence of spacetime itself, or a ``bubble of nothing''.  
This connection has been made precise in
various contexts \cite{Horowitz:2005vp,Hirano:2005sg,Headrick:2004hz,Gutperle:2002bp,Emparan:2001gm,Costa:2000nw,Adams:2005rb,Aharony:2002cx,zamolunpub}. 

In these solutions, the tachyon is homogeneous in the $X\ll 2, \ldots , X\ll{D-1}$
directions.  One can also consider situations where the tachyon has some
oscillatory dependence on other coordinates.  For instance,
we can consider a superposition of perturbations
\bbb
\ct (X) = \m\ll 0\sqd \cc \exp{\b X\uu +} - \m\ll k\sqd\cc
 \cos(k X\ll 2) \exp{\b\ll k  X\uu +} \ ,
\eee
with 
\bbb
q \b\ll k  = \sqrt{2} \lrdd {2\over\apr} - \hh k\sqd \rrdd\ .
\eee
The tachyon couples to the string worldsheet as a potential:
\bbb
\Delta {\cal L} = -{1\over{2\pi}} :{\cal T}(X) :\ .
\eee
The theory has a vacuum at $X\ll 2 = 0$,
with potential
\bbb
\ct = \ct\ll 0 + \hh k\sqd
\m\ll k \sqd \cc\exp{\b X\uu +}\cc
: X\ll 2 \sqd :+ O(k\uu 4 X\ll 2\uu 4)\ ,
\eee
where we have defined
\bbb
\ct\ll 0 \equiv \m\ll 0 \sqd \cc \exp{\b X\uu + } - 
\m\ll k \sqd  \cc \exp{\b\ll k  X\uu +} \ .
\eee
The theory simplifies considerably when the wavelength $k\uu{-1}$ of
the tachyon is long compared to the string scale.  Taking 
$k\to 0$
with $\apr k\sqd \m\ll k \sqd \equiv \m \sqd$
and $\m\uu{\prime 2} 
\equiv \m\ll 0\sqd - \m\ll k
\sqd$ held fixed,
the growth
rate $\b\ll k $ approaches $\b$,
and the $X\ll 2\uu 4$ and higher terms in the potential
vanish.  The tachyon becomes simply:
\bbb
\ct(X\uu +, X\ll 2)
 = + {{\m \sqd}\over{2 \cc \apr}}
 ~\exp{\b X\uu +} ~:X\ll 2\sqd:
+ \ct\ll 0 (X\uu +)\ ,
\xxx
\ct\ll 0 (X\uu +) =  {{\m\sqd\cc X\uu +}\over
{\apr\cc q\cc\sqrt{2}}} \cc
\exp{\b X\uu +} + 
\m\uu{\prime 2}\cc
\exp{\b X\uu +} \ .
\een{newlabel}

Intuitively, the physics of the solution seems
rather simple.  At $X\uu + \to - \infty$, the
string is free to propagate in all $D-1$ spatial
dimensions.  As the string reaches a regime where
$X\uu 0 \simeq - X\uu 1$, there is a
worldsheet potential
that confines the string to the origin of the
$X\ll 2$ direction.  In the region
$X\uu + \to +\infty$, strings propagate in a
lower number of spatial dimensions altogether (specifically, $D-2$).
Strings which continue to oscillate in the $X\ll 2$
direction will be expelled from the region of large
tachyon condensate and pushed along the domain
boundary $X\ll 1 \simeq - X\uu 0$ at the speed of
light.
This is the essential physics of
dynamical spacetime dimension change, at the
level of semiclassical analysis on the string 
worldsheet.

In this interpretation, 
the meaning of the operator
$\ct\ll 0 (X\uu +)$ is mostly, though not yet
entirely self-evident.
The most straightforward 
interpretation would be that $\ct\ll 0$
should be thought of
as representing tachyon condensation along
a lightlike direction in the lower-dimensional
string theory, in $D-1$ spacetime dimensions.
The parameter $\m\pr\sqd$ then becomes the amplitude
of a mode $\exp{\b X\uu +}$
of the tachyon, which one can fine-tune to
zero by setting $\m\uu{\prime 2}$ 
to vanish.\footnote{At the quantum level, ${\mu'}^2$ must be fine-tuned
to a nonzero value to cancel an additive regulator-dependent contribution 
from vacuum fluctuations of the $X\ll 2$ field.}

Less transparent is the meaning of the
$X\uu + \cc \exp{\b X\uu +}$ term in $\ct\ll 0$.
One cannot fine tune its
coefficient by hand, since
neither $X\uu + \exp{\b X\uu +}$
nor $\exp{\b X\uu +} :X\ll 2\sqd :$
is by itself an operator
of definite weight $(1,1)$.  Only the combination
that appears in $\ct(X\uu +, X\ll 2)$ is an
operator of eigenvalue one under $L\ll 0$ and
$\tilde{L}\ll 0$.  This seems to leave
the effective $(D-1)$-dimensional string theory
with a tachyon condensate, growing as $X\uu + \cc
\exp{\b X\uu +}$, which cannot be fine-tuned away.
Without removing the $X\uu + \cc \exp{\b X\uu +}$
term, we are left without an interpretation
of our model as interpolating between two string
theories with zero tachyon at $X\uu + = \pm\infty$ in different
numbers of dimensions.

We shall see presently that this term will be precisely
canceled by the quantum effective potential
generated upon integrating out the $X\ll 2$ field.
In retrospect, this is inevitable.  The effective
tachyon in the lower-dimensional string theory
must couple to the string worldsheet as an
operator of definite weight: this allows the
term $\m\uu{\prime 2} \exp{\b X^+}$ appearing in $\ct_0$, 
but forbids any coupling that scales as $X\uu + \cc \exp{\b
X\uu +}$ in the effective theory.

\subsection{Classical solutions of the worldsheet CFT}

The string-frame metric is flat, and so the kinetic term for the fields is
\be
{\cal L}\ll{\rm kin} &=&
{1\over{2\pi\apr}} 
\Bigl[ -
(\pp\ll {\s\uu 0} X\uu +)(\pp\ll {\s\uu 0 }X\uu -)
+ (\pp\ll {\s\uu 1} X\uu +)(\pp\ll {\s\uu 1} X\uu -) 
\nn\\ && \nn\\
&&	+ \hh (\pp\ll {\s\uu 0} X\uu i)  (\pp\ll {\s\uu 0} X\uu i)  
	-
	\hh (\pp\ll {\s\uu 1} X\uu i)  (\pp\ll {\s\uu 1} X\uu i)  
	\Bigr]\ .
\ee
Taking $\s\uu\pm \equiv 
- \s\uu 0 \pm \s\uu 1,~ \pp_\pm = \frac{1}{2}(-\pp\ll{\s^0} \pm \pp\ll{\s\uu 1})$, 
the equations of motion are
\bbb
\pp\ll + \pp\ll -  X\uu + = + 
{{\apr}\over 4} \pp\ll{X\uu -}\ct = 0\ ,
\xxx
\pp\ll + \pp\ll - X\ll 2 
= - {{\apr}
\over 4} \pp\ll{X\ll 2}\ct = - {1\over 4} \m \sqd
\exp{\b X\uu +}~X\ll 2\ ,
\xxx
\pp\ll + \pp\ll - X\uu - 
=  + {{\apr}\over 4} \pp\ll{X\uu +}\ct = + {\b\over 8}
\m\sqd \exp{\b X\uu +}~X\ll 2\sqd + {1\over 4}
\ct\ll {0,X\uu +}\ ,
\xxx
\pp\ll + \pp\ll - X\uu i = 0,~~~~~  \qquad i = 2,\cdots , D-1\ .
\eee
Treating $X\uu +$ as fixed, the effect of
the interaction term on $X\ll 2$ is to give
the field a physical mass of magnitude $M(X\uu +)
\equiv \m \cc \exp{\hh \b X\uu +}$.

The equations of motion can be solved in
full generality on a noncompact worldsheet. 
Schematically, the general solution for $X\uu +$ can be expressed as
\bbb
X\uu + = f\ll + (\s\uu +) + f\ll - (\s\uu -)\ .
\een{littlefdef}
Given $f\ll{\pm}$, antiderivatives may then be defined as:
\bbb
\frac{d}{d\s^\pm} {\bf F} \ll\pm (\s\uu\pm)  \equiv \exp{\b f\ll\pm (\s\uu\pm)} \ .
\een{bigfdef}
The general solution to the equation of motion for $X\ll 2$ 
is thus a linear combination of solutions of the form
\bbb
X\ll 2 = \sum\ll \o A\ll \o  \expr {{i\over 2} \o {\bf F}\ll + (\s\uu +)
+ {i\over 2}  {{\m\sqd}\over \o}  {\bf F}\ll - (\s\uu -)}\ ,
\eee
for an arbitrary set of amplitudes $A\ll \o$.

Since the field equations are
nonlinear, it is more complicated to find the general solution in
finite volume, and we will not do so here.
However, we can understand the finite-volume theory
by looking at pointlike solutions in which the coordinates
$X\uu{\pm},~X\ll 2$ are independent of $\s\uu 1$.  Then $X\uu +$ is of
the form
\bbb
X\uu + = \apr p\uu + (\s\uu 0 - \s\uu 0\ll {(0)})  \ ,
\eee
so the equation of motion for $X\ll 2$ becomes
\bbb
\Xdd\ll 2 = - \m\sqd \exp{\b \apr p\uu + (\s\uu 0 - \s\uu 0\ll{(0)})} X\ll 2\ .
\eee
The solution for $X\ll 2$ can be expressed in terms of Bessel functions:
\bbb
X\ll 2 &=& A\, J\ll 0 \lrdd {{2 \m ~\exp{\hh \b \apr p\uu + (\s\uu 0 -
\s\uu 0\ll{(0)}) }}\over{\b\apr p\uu +}} \rrdd
\nn\\
&&
+  B\, Y\ll 0 \lrdd {{2 \m ~\exp{\hh \b \apr p\uu + (\s\uu 0 -
\s\uu 0\ll{(0)}) }}\over{\b\apr p\uu +}} \rrdd\ .
\eee
In particular, $J\ll 0,~Y\ll 0$ are respectively Bessel functions of the first and second kind, 
and $A,~B$ are constants of motion.
The particle behaves precisely as a harmonic oscillator with
a time-dependent frequency
\bbb
\o(\s\uu 0 ) \equiv 
M(\s\uu 0) =  
\m\cc\exp{\hh \b\apr p\uu + (\s\uu 0 - \s\uu 0\ll {(0)})}\ .
\eee
As predicted by the adiabatic theorem, energy in the oscillator grows in
proportion to $\o$.  Although we are only solving the equation
classically, we write 
\be
\hh \Xd\ll 2 \sqd + \hh M\sqd (\s\uu 0) X\ll 2 \sqd 
\equiv {{\hbar \apr}\over{\rws}}
 \cc \numosc (\s\uu 0) M (\s\uu 0)\ ,
\ee  
where $\rws $ is the worldsheet radius and $\numosc (\s\uu 0)$ counts the 
number of oscillator excitations.
It is clear that $\numosc(\s\uu 0)$ asymptotes to a constant $\numosc\fin$ in the
limit $\s\uu 0\to \infty$.  By the virial theorem, the potential and
kinetic energies separately approach $\hh \hbar \apr\cc \numosc\uu{\rm (fin.)}
 M (\s\uu 0)$,
on average.  So, in particular, we find that
\bbb
\hh\m\sqd \exp{\b X\uu +(\s) } X\ll 2 \sqd(\s) \to 
{{\m\apr}\over {2\rws}} \cc \hbar \numosc\fin
\exp{\hh \b X\uu + (\s)}  \ ,
\eee
as $\s\to\infty$.

The equation of motion for $X\uu -$ appears as
\bbb
\ddot X\uu - = \hh \b\m\sqd \exp{\b X\uu +} 
X\ll 2 \sqd\ ,
\eee
which means that
\bbb
X\uu - \simeq \hh {{\m\sqd}\over{\b p\uu{+2} \apr\sqd}} \exp{\b X\uu +} X\ll 2\sqd 
\simeq  {{\m\sqd \numosc\fin \hbar}\over{2\b p\uu{+2} \apr \rws }} \exp{\hh \b X\uu +}  \ ,
\eee
as $\s\uu 0\to \infty$.

We can interpret this as follows.
Once the particle meets the bubble wall, it accelerates rapidly to the
speed of light and is pushed along the wall for all time, unless 
$\numosc\uu{{\rm(fin.)}} = 0$.
The conclusion is intuitively obvious: strings with excited degrees of freedom 
corresponding to the $X\ll 2$ direction are forbidden energetically from
entering the interior of the bubble.  
%
Allowing $\s\uu 1$-dependent modes $\exp{i n\s\uu 1}$ of the $X\ll 2$
field to be excited does not change this picture.  Each has a
time-dependent frequency
\bbb
\o\ll n(\s\uu 0) \equiv \lrdd M\sqd(\s\uu 0) + {{n\sqd}\over{\rws\sqd}} \rrdd \uu \hh\ ,
\eee
and an energy that asymptotes to a constant multiple of
$\o\ll n(\s\uu 0)$:
\bbb
E\ll n \equiv N\ll n (\s\uu 0) \o\ll n (\s\uu 0) \ ,
\xxx
\numosc\ll n(\s\uu 0) \to  N\ll n\uu{\rm (fin.)} \ ,
\eee
as $\s\uu 0\to\infty$.
Only when $\numosc\fin\ll n = 0$ for all modes
is it possible for the particle to enter the region $X\uu 1 \gtrsim  -X\uu 0 $.

\subsection{Physics at large $X\uu +$}
At late times, the space of solutions to the equations of
motion bifurcates into two sectors.
First, there is a sector
of states that are accelerated along the bubble wall to near the speed of
light.  These states are pinned to the wall of the bubble, and they
have an energy that increases quickly as a
function of time.  The dynamics of these states is
essentially the same as the dynamics of the final
states in the solution described in \cite{previous}.
These are the string states with 
modes of $X\ll 2$ excited. At large
$\s\uu 0$, the frequencies of these
modes increase exponentially with time, so the condition of the
adiabatic theorem is satisfied:
\bbb
{1\over{\o\ll n\sqd}} {{d\o\ll n}\over{d\s\uu 0}}  \sim \hh \b \apr p\uu + \cc \exp
{- \hh \b p\uu + \apr \s\uu 0} \to 0,~~~~~ {\rm as}~~ \s\uu 0 \to + \infty\ .
\eee
In the limit $\s\uu 0 \to +\infty$,
all modes of $X\ll 2$ are frozen into their excited states, with fixed energies in
units of $\o\ll n(\s\uu 0)$.
Recalling that $X\uu +$ is proportional to worldsheet time, with
positive coefficient of proportionality $\b p\uu + \apr$, we learn that modes
pushed along the bubble wall will lose their ability cross through to the
interior, as energy becomes locked permanently into their $X\ll 2$ modes.

Second, we have states for which
$\numosc\ll n (\s\uu 0) \to 0$ for all oscillators
as $\s\uu 0 \to +\infty$.  For these
states, the modes of $X\ll 2$ are in
their ground states, and $X\ll 2$ can be integrated out.
States with all $X\ll 2$ oscillators in their ground states
have vanishing classical vacuum energy in the $X\ll 2$
sector.  Quantum mechanically, their energy is $\hh \hbar
\omega\ll n (\s\uu 0)$ for each oscillator.  The total ground
state energy of the oscillators as a function of time is then
\bbb
E\ll 0(M(\s\uu 0) ; \L; \rws) &=&
\hh\hbar \sum\ll{n = -\infty}\uu\infty
\lrdd {{n\sqd}\over{\rws\sqd}} + M\sqd (\s\uu 0) \rrdd\uu{\hh}~\exp
{-  n\sqd /\rws\sqd  \L\sqd} \ 
\nn\\&&\nn\\
&\equiv& 2\pi \rws \vwsol (M;\L) + O\lrdd {1\over{\rws}} \rrdd\ , 
\eee
where $\L$ is a mass scale introduced to regulate the
sum, and $2\pi \rws$ is the
spatial volume of the worldsheet.

As $\L\to \infty$ there are various infinite,
$\L$-dependent terms in the quantum vacuum energy
$\vwsol(M;\L)$.
Part of this vacuum energy is present in
the $M\to 0$ limit, and this piece of the
energy depends on the
structure of the ultraviolet regulator.  It
should be subtracted with a local counterterm
as part of the definition of the worldsheet
path integral with vanishing tachyon:
\be
E\ll 0\uu{{\rm renormalized}} &\to & E\ll 0\uu{{\rm bare}} 
- 2\pi \rws \vwsol (0;\L)\ ,
\nn \\ && \nn \\
\vwsol (M;\L) &\equiv & 
{{\hbar}\over{2\pi \rws}} \int\ll 0\uu\infty dn~
\exp{- n\sqd / \rws\sqd\L\sqd}
~
\lrdd {{n\sqd}\over{\rws\sqd}} + M\sqd \rrdd\uu{\hh}\ ,
\nn \\ && \nn \\
&=& {{\hbar}\over{2\pi}}  \L\sqd \int\ll 0\uu\infty dy~
\exp{- y\sqd}
~
\lrdd y\sqd + {{M\sqd}\over{\L\sqd}} \rrdd\uu{\hh}\ .
\ee
After subtracting $\vwsol$ at $M=0$, the remainder of
the one-loop effective potential is
\be
\vwsrem(M;\L) &\equiv& \vwsol (M;\L) - \vwsol(0;\L) 
\nn\\ && \nn\\
&=& {\hbar\over{2\pi}} \int\ll 0\uu\infty {{M\sqd ~dy}\over{
y + \sqrt{y\sqd + {{M\sqd}\over{\L\sqd}}}}} \exp{- y\sqd}\ ,
\nn \\ && \nn \\
&=&  
{\hbar\over{2\pi}} M\sqd \lrdd a\ll 1 + a\ll 2 \ln \lrdd{{M\sqd}
\over{\L\sqd}}  \rrdd  \rrdd + O\lrdd {{M\uu 4}\over{\L\sqd}}
\ln \lrdd {{M\sqd}\over{\L\sqd}} \rrdd \rrdd\ ,
\een{a1piece}
where we have defined the following quantities:
\bbb
a\ll 1 &\equiv& 
\int\ll 0\uu\infty {{ ~dy}\over{
y + \sqrt{y\sqd + 1 }}} \exp{- y\sqd}\ ,
\nn\\&&\nn\\
a\ll 2 &\equiv& 
 - {1\over 4} 
\ .
\eee
The remainder $\vwsrem(M;\L)$ 
vanishes for $X\uu + \to -\infty$, and becomes large as $X\uu +
\to\infty$.  
It consists of a term proportional
to $M\sqd \equiv \m\sqd \cc\exp{\b X\uu + }$, and a term
proportional to $M\sqd \ln(M\sqd / \m\sqd)
= \m\sqd \b X\uu +\cc \exp{\b X\uu + }$.
As noted earlier, the first 
is a conformal perturbation, so it can be
subtracted.  
This may be thought of as
fine-tuning an initial condition so that the
bosonic string tachyon of the $D-1$ dimensional
string theory vanishes in the far future.   In
practice, performing this subtraction amounts to
adjusting the coefficient $\m\uu{\prime 2}$ in expression~\rr{newlabel} 
to cancel the term proportional to $a\ll 1$
in Eqn.~\rr{a1piece} above.  

For definiteness, we chose
to regulate the sum with the factor $\exp
{-  n\sqd /\rws\sqd  \L\sqd}$, but the results would be
the same had we replaced $\exp
{-  n\sqd /\rws\sqd  \L\sqd}$ with ${\bf B} (n\sqd /\rws\sqd 
\L\sqd)$, where ${\bf B}(s)$ is a smooth
bump function of arbitrary shape,
which satisfies $B(0) = 1$ and approaches zero with all
its derivatives when its argument goes to infinity.  The
only difference would have been a different value for the
coefficient $a\ll 1$ of $M\sqd$.  The coefficient 
$a\ll 2$ is equal to $-{1\over 4}$ for any form of the
regulator.

We are left with the $M\sqd \cc\ln (M\sqd / \m\sqd)$
term, which contributes
\bbb
\Delta \vwsrem =
- {{\m\sqd}\over{8\pi}}~\b X\uu + \cc \exp{\b X\uu +}
\een{quantumnonconformal}
to the vacuum energy.  There is a term of
the same form that we were forced to add
to the classical action
to make the tachyon perturbation marginal:
\bbb
\Delta \vwscl = {1\over{2\pi}} (\Delta \ct ) =
{1\over{2\pi}}
{{\m\sqd X\uu +}\over{\apr q \sqrt{2}}}~\exp{\b
X\uu +}\ .
\eee
Using the relation $\b q = {{2\sqrt{2}}\over\apr}$,
we find that this addition is equal to 
\bbb
\Delta \vwscl  =
+ {{\m\sqd}\over{8\pi}} 
(\b X\uu +)
\exp{\b X\uu +}\ ,
\een{classicalnonconformal}
which precisely cancels the term $\Delta \vwsrem$
in Eqn.~\rr{quantumnonconformal}.  


The extensive piece of the vacuum energy,
scaling like $\rws\uu {+1}$, vanishes
between the tree-level terms $(\b X\uu +)\cc\exp{\b X\uu +}$
and $\exp{\b X\uu +}$, the one-loop effective terms,
and the counterterms.
The next most
important term at large volume, scaling like $\rws\uu {-1}$,
is the Casimir energy, which is related to the central charge.

The Casimir energy  $E\ll 0 (M;\L;\rws)
- 2\pi \rws \vwsol(M;\L)$ in the $X\ll 2$ sector changes as
we go from $X\uu + = -\infty$ to $X\uu + = + \infty$.  
At $X\uu + = - \infty$, the mass of the $X\ll 2$ field
vanishes, and the Casimir energy in the $X\ll 2$
sector is $-{1/{12 \rws}}$,
as usual for a {\it massless} field.  As
$X\uu + \to + \infty$, the Casimir energy approaches zero
as $\exp{- M\rws}$, which would be expected 
for a {\it massive} field.  In particular, the
$1/ \rws$ term in the energy vanishes at $X\uu + \to\infty$.
%
%
We can infer from this that the central charge in the $X\ll 2$
sector changes between $X\uu + = -\infty$ and $X\uu + =
+ \infty$.  
In the full theory of the $X\uu \pm,~ X\ll 2$ fields,
the total central charge is independent
of $X\uu +$, as it must be for
consistency of the CFT.  
We will now discuss the
quantum properties of the
theory and see directly how the 
metric and dilaton are renormalized
by a loop diagram, giving
rise to the necessary adjustment
of the central charge.

\subsection{Structure of quantum corrections in the worldsheet CFT}
The properties that render the theory exactly
solvable at the classical level also give the
quantum theory very special features.  The most
striking aspect of the quantum worldsheet theory
in the dimension-changing
transition is that the worldsheet CFT is
not free, but its perturbation series is
simple enough that all quantities can be
computed exactly.  In particular, all connected
correlators of free fields have perturbation
expansions that terminate at one-loop order.

This can be seen directly from the structure of
Feynman diagrams in the $2D$ theory.\footnote{Very similar arguments were 
used to constrain quantum corrections in \cite{previous, atnulltach1, atnulltach2}.}  
By the underlying target space Lorentz invariance of the
theory on a flat worldsheet, the propagator
$\left\langle X\uu\m X\uu\n \right \rangle$
is proportional to $G\uu{\m\n}$, so it
only connects ``$+$'' fields to ``$-$'' fields.  
Denoting the massive field $X\ll 2$ with a solid line 
and the $X\uu\pm$ fields by dashed lines,
we may therefore draw massless propagators as oriented, 
with arrows pointing from $+$ to $-$.

Since the vertices depend only on $X\uu +$ and not on
$X\uu -$, the interaction vertices have only outgoing, 
and no incoming, dashed lines. 
It follows immediately that no two vertices can
be connected with a dashed line.  Furthermore, 
every vertex has exactly zero or two solid lines
passing through it, and we deduce that every connected
Feynman diagram must have either zero or one loops.  To see
this, ignore all dashed lines in the Feynman diagram:
this gives a collection of solid line segments
and solid loops.  If there is more than one
solid piece, then the full diagram (including
dashed lines) must be disconnected, since one
can never connect two disconnected vertices using 
dashed lines.

\begin{figure}[htb]
\begin{center}
\includegraphics[width=6.0in,height=1.4in,angle=0]{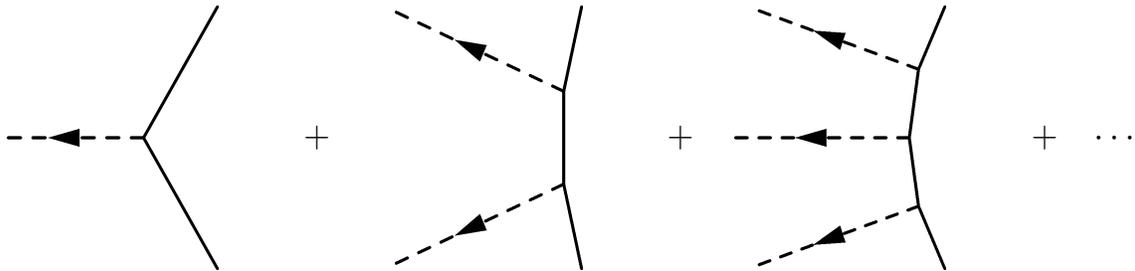}
\caption{Tree diagrams containing multiple vertices
have the structure of an ordered
sequence of vertices with a solid line passing
through.  Here, a single massless leg emerges from each vertex.}
\label{orderlegs}
\end{center}
\end{figure}

We conclude that every
connected tree diagram with multiple vertices
has the structure of an ordered
sequence of vertices with a solid line passing
through, and arbitrary numbers of dashed lines
emanating from each vertex.   This is depicted in Fig.~\ref{orderlegs},
in the particular case where one massless line emerges from each 
individual vertex.

\ \\

\begin{figure}[htb]
\begin{center}
\includegraphics[width=6.0in,height=1.4in,angle=0]{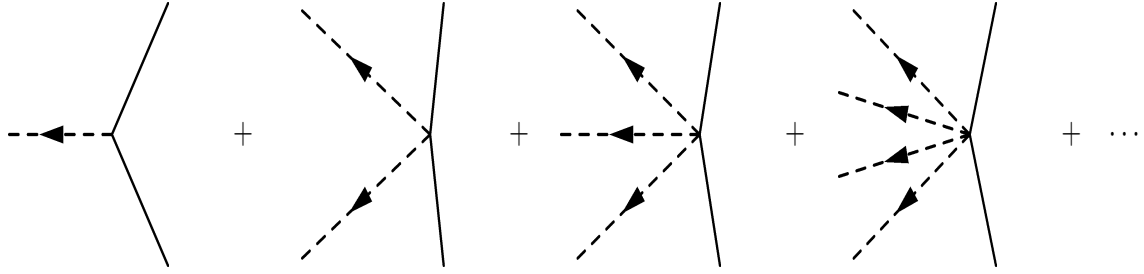}
\caption{One-vertex tree diagrams composed of a number of outgoing, massless
legs and exactly two massive legs. }
\label{twolegs}
\end{center}
\end{figure}

\begin{figure}[htb]
\begin{center}
\includegraphics[width=1.15in,height=1.8in,angle=0]{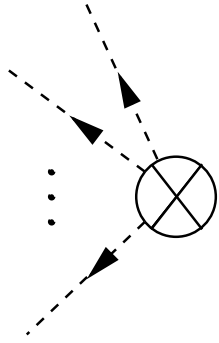}
\caption{The tree-level potential and the counterterms include vertices depending
only on $X^+$ (depicted here by dashed, oriented lines).}
\label{counterterm}
\end{center}
\end{figure}

Fundamental vertices representing the
counterterms and classical potential $\exp{\b X\uu +}$, 
$(\b X\uu +) \exp {\b X\uu +}$ and $X\ll 2\sqd\cc \exp{\b X\uu +}$,
have arbitrary numbers of
dashed lines emanating from them, and two or zero
solid lines. These diagrams are shown
in Fig.~\ref{twolegs} and \ref{counterterm}.

\ \\

\ \\

\begin{figure}[htb]
\begin{center}
\includegraphics[width=6.0in,height=2.8in,angle=0]{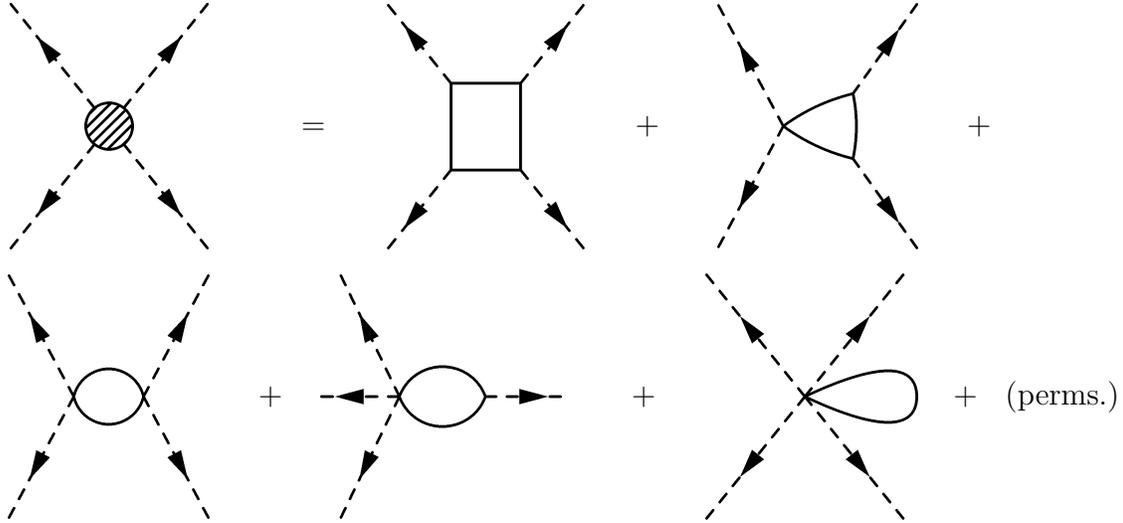}
\caption{The four-point Feynman diagram, exact at one-loop order.}
\label{fig4}
\end{center}
\end{figure}

The connected loop diagrams consist strictly of a closed solid line
with dashed lines emanating from an arbitrary 
number of points on the solid line.  This is depicted for the 
four-point interaction in Fig.~\ref{fig4}.
This classification exhausts the set of
connected Feynman diagrams in the theory.
In particular, {\it every} connected correlator is exact
at one loop.

\ \\

\subsection{Dynamical readjustment
of the metric and dilaton gradient}
The one-loop diagrams can be thought of
as a set of effective vertices for $X\uu +$, 
associated with integrating out the
massive field $X\ll 2$.
The operators generated in this fashion 
redefine the background fields in the target
space of the lower-dimensional string theory
at large $X\uu +$.  Most of these 
fields vanish in the limit $X\uu + \to \infty$.  
To see this, note that an operator with
canonical weight $(h,h)$ can only appear dressed
with $M\uu{1-h}$.  For instance, the
operator
$(\pp\ll - X\uu +)
(\pp\ll - \sqd X\uu +)(\pp\ll -\uu 3 X\uu +)
(\pp\ll + X\uu +)
(\pp\ll + \sqd X\uu +)(\pp\ll +\uu 3 X\uu +)$
can only acquire the dressing $\m\uu{-10}\cc\exp{- 5 \b X\uu +}$.
Since all undressed operators that appear have
integer weight, the only operator
with increasing dependence on $X\uu +$ is the
identity, dressed with $\exp{\b X\uu +}$.  However, 
this is just the coupling of the lower-dimensional 
effective tachyon, which we have fine-tuned to zero by 
adjusting $\m\uu{\prime 2}$.  

Dressed operators of weight $(2,2)$ and higher die
off with increasing $X\uu +$.  Therefore, the
only $2D$ couplings on which the loop diagrams
have any effect in the $X\uu +\to \infty$ limit
are the string frame metric $G\ll{\m\n}$,
which couples to $(\pp\ll + X\uu \m )(\pp\ll -
X\uu \n)$, and the dilaton $\Phi$, which couples
to the worldsheet Ricci scalar.  (Worldsheet
parity symmetry, among other constraints,
forbids a renormalized background for the
$B\ll{\m\n}$ field in this solution).

It is not difficult to integrate out $X\ll 2$
and extract the one-loop shifts in these two
couplings.  
Integrating out a real scalar with mass $M$ contributes to the
effective dilaton (see Ref.~\cite{hellerman2} for further details) 
by an amount
\bbb
\Delta \Phi = + {1\over 6} \ln \lrdd {M\over{\tilde{\m}}} \rrdd\ ,
\een{dilrenormone}
where $\tilde{\m}$ is an arbitrary mass scale 
involved in the definition of the path integral measure.  Since
$M\equiv \m\cc \exp{\hh \b X\uu +}$, this gives
\bbb
\Delta \Phi = \lrdd {\rm const.}  \rrdd + 
{\b\over {12}}  X\uu +\ .
\een{dilrenormtwo}

\begin{figure}[htb]
\begin{center}
\includegraphics[width=3.0in,height=1.5in,angle=0]{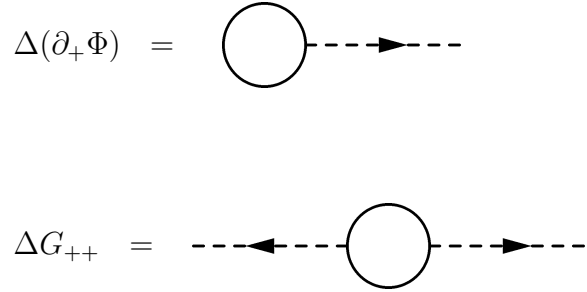}
\caption{Diagrams contributing to the nonvanishing 
renormalizations of the dilaton and metric, 
$\Delta(\partial_+ \Phi)$ and $\Delta G\ll{++}$.}
\label{renorm}
\end{center}
\end{figure}

There is also a nonzero
renormalization of the string-frame metric.
For a mass term $\hh M\sqd (X) ~X\ll 2\sqd$, where 
$M$ depends arbitrarily on all coordinates other
than $X\ll 2$, the metric $G\ll{\m\n}$ is renormalized
\cite{hellerman2} by an amount
\bbb
\Delta G\ll{\m\n} = +{{\apr}\over 6} {{\pp\ll{\m}M  \pp\ll\n M}
\over{M\sqd}} \ . 
\een{metrenormone}
For $M = \m \cc\exp{\hh \b X\uu +}$, this gives
\bbb
\Delta G\ll{++} = +{{\apr\b\sqd }\over {24}}\ ,
\een{metrenormtwo}
with all other components unrenormalized.
The renormalized upper-indexed metric then reads
\bbb
G\uu{--} = - {{\apr\b\sqd }\over {24}}\ ,
\eee
with all other upper-index metric components unrenormalized.
The renormalization of the dilaton and metric
is depicted diagrammatically in Fig.~\ref{renorm}.

\ \\ 

Let us now check the linear dilaton contribution to the
central charge in the $X\uu + \to \infty$ limit. 
From the above analysis, we find a renormalized dilaton gradient
$\hat{V}\ll\m \equiv V\ll\m + (\Delta V)\ll\m $
and string frame metric $\hat{G}\ll{\m\n} \equiv 
G\ll{\m\n} + \Delta G\ll{\m\n}$ given by
\bbb
\hat{V} \ll - = V\ll - = - {q\over{\sqrt{2}}}\ ,
\xxx
\hat{V} \ll +  = - {q\over{\sqrt{2}}}
+ {\b\over{12}} \ ,
\xxx
\hat{G}\uu {+-} = \hat{G}\uu{- +} = -1\ ,
\qquad 
\hat{G} \uu{--} = - {{\apr\b\sqd }\over {24}}\ ,
\qquad
\hat{G}\uu{++} = 0\ .
\eee
In the limit $X\uu +\to +\infty$,
we therefore recover a contribution to the central charge
from the renormalization of the linear dilaton equal to
\bbb
c\uu{\rm dilaton} = 6\apr \hat{G}\uu{\m\n}
\hat{V} \ll\m \hat{V}\ll \n = - 6\apr q\sqd 
+ {{q\b\apr}\over{\sqrt{2}}} - {{\apr\sqd q\sqd\b\sqd}
\over 8}\ .
\eee
Using $q\sqd = (D - 26)/6\apr$ and $q\b = {{2\sqrt{2}}\over{\apr}}$,
we find, in the $X\uu + \to \infty$ limit,
\bbb
c\uu{\rm dilaton} = 6\apr \hat{G}\uu{\m\n}
\hat{V} \ll\m \hat{V}\ll \n = - (D - 26) + 1 \ .
\eee
The central charge contribution from $X\ll 2$ decreases by
one unit in this limit, so the
total central charge remains the same.  Instead
of an overall net change, units of central charge contribution
have merely been transferred from those of the free scalar $X\ll 2$ to 
those coming from the strength of the dilaton gradient.  
Henceforth, we refer to this effect as {\it central charge transfer.}

The central charge transfer mechanism works equally well
when the tachyon has a quadratic minimum in several dimensions
$X\ll 2,\cdots, X\ll{n+1}$, rather than just one.  
In that case, the worldsheet
Lagrangian is deformed by a term $\Delta {\cal L} = - {1\over{2\pi}} \ct (X)$, with
\bbb
\ct (X) = 
  {{\m \sqd}\over{2 \cc \apr}} 
 ~\exp{\b X\uu +} ~\sum\ll{i = 2}\uu {n + 1} \cc:X\ll i\sqd:
+ \ct\ll 0 (X\uu +)\ ,
\xxx
\ct\ll 0 (X\uu +) =  {{n\cc\m\sqd\cc X\uu +}\over
{\apr\cc q\cc\sqrt{2}}} \cc
\exp{\b X\uu +} + 
\m\uu{\prime 2}\cc
\exp{\b X\uu +} \ .
\eee
The renormalizations of $\hat{G}\uu{\m\n}$ and
$\hat{V}\ll \m$ receive contributions from $n$ real scalars,
rather than just one:
\bbb
\hat{V} \ll - = - { q\over{\sqrt{2}}}\ ,
\qquad 
\hat{V} \ll + = - {q\over{\sqrt{2}}}
+ {{n \b}\over{12}} \ ,
\xxx
\hat{G}\uu {+-} = \hat{G}\uu{- +} = -1\ ,
\qquad
\hat{G} \uu{--} = - {{n \apr\b\sqd }\over {24}} \ ,
\qquad
\hat{G}\uu{++} = 0\ .
\eee
The dilaton contribution to the central charge in 
the $X\uu + \to \infty$ limit becomes
\bbb
c\uu{\rm dilaton} = 6\apr \hat{G}\uu{\m\n}
\hat{V} \ll\m \hat{V}\ll \n = - 6\apr q\sqd 
+ {{n q\b\apr}\over{\sqrt{2}}} - {{n \apr\sqd q\sqd\b\sqd}\over 8}
= - (D - 26) + n\ ,
\eee
which compensates the loss of $n$ units of
central charge carried by the $n$ real scalars $X\ll 2,
\cdots , X\ll {n + 1}$.

The central charge transfer described here is the
same as the mechanism studied in \cite{hellerman2}.  
In the model studied in \cite{hellerman2}, the tachyon condenses along
a timelike rather than lightlike direction.  As a result,
there are Feynman diagrams involving loops of the $X\uu 0$
field that make contributions to the renormalized
metric and dilaton, suppressed by powers of
$ n / (D - D_{\rm crit})$.  The CFT describing timelike tachyon
condensation can only be treated semiclassically
if $D >> D_{\rm crit}$ and $n$ is small, so the description
\cite{hellerman2} can never be perturbative when
describing a return to the critical vacuum.
In the present model,
the one-loop renormalizations are exact, and no large-$D$ limit
need be taken.   We can therefore study transitions in which
the number of spacetime dimensions at $X\uu + \to \infty$
is equal to the critical dimension or less!

To make the final number of dimensions equal to the
critical dimension $D\ll{\rm crit} = 26$, choose
$n = D - 26$. 
The final
dilaton gradient is nonvanishing, but it is null
with respect to the final string-frame metric.
Furthermore, the product $\hat{G}\uu{\m\n} \hat{V}\ll\m {{\pp\ll\n \ct}/{\ct}}$
is positive, signaling a dilaton rolling to
infinitely weak coupling in the future of the final,
critical vacuum.\footnote{Since the changes
in the metric and dilaton are discrete rather
than continuous, it is not completely obvious how to identify the future
in the lower-dimensional string theory. The coordinate $X\uu +$ can
be read off from the decay of the background massive modes: massive
modes decay with {\it increasing} $X^+$.   
Since the gradient of
$X\uu +$ has a positive inner product with $\hat{V}\ll\m$ when computed with
respect to $\hat{G}\uu{\m\n}$, the weak coupling direction
is identified as a future-oriented lightlike vector.}

In the remainder of this paper we study
several models describing tachyon condensation
along lightlike directions in type 0/type II and heterotic string 
theories.  In each of the
models we study, amplitudes receive
quantum corrections from Feynman diagrams
with at most one loop, and central charge is transferred from
free fields to the strength of the dilaton gradient in
such a way that the theories at $X\uu + = \pm\infty$ have
the same total central charge, despite living in different numbers 
of spacetime dimensions.


%
%
\section{Dimension-changing transitions in heterotic string theory}
%
We will now describe exact classical solutions of 
heterotic string theory in which
the total number of spacetime dimensions decreases by $n$.
In the case $n = D - 10$, the final vacuum has the
critical number of dimensions $D\ll{\rm crit} = 10$, with
flat string-frame
metric, and lightlike linear dilaton rolling to weak coupling
in the future.  The particular heterotic theory we 
study has gauge symmetry $SO(22 + D)$ and a tachyon $\ct \uu a$
transforming in the fundamental representation.  This
is the second of the two unstable heterotic theories described in 
\cite{hellerman1},\footnote{The solutions described in this paper go through equally well
for the first of the two theories in \cite{hellerman1}, as long as $n \leq D - 10$.
We focus on type HO$^{+/}$ because its GSO projection is simpler.}
named type HO$^{+/}$.  The tachyon
in the type 0 theory has a single real component
$\ct$, as always.

If the $D$-dimensional theory at $X\uu + = -\infty$ has
its spacelike dimensions in the form of the maximally
Poincare-invariant $D-1$ dimensional $\IR\uu {D-1}$, 
then the final theory is type HO$^{+/}$ in
$D-n$ dimensions, with spatial slices of the form
$\IR\uu{D-1-n}$.

Alternately, the spatial slices of the initial theory
can be given certain orbifold singularities on
which light spacetime fermions can live.  The lower-dimensional
theory at $X\uu + = + \infty$ then has spacetime fermions as
well.  In the case where the final dimension is critical, we
can reach the spacetime-supersymmetric $SO(32)$ heterotic theory.  

In all cases, the readjustment of the dilaton and
the string-frame metric comes from a one-loop calculation on the
worldsheet, just as in the dimension-changing transition we
studied in the bosonic string in the previous section.

\subsection{Description of type HO$^{+/}$ string theory in $D > 10$}
In heterotic string theory, the local worldsheet gauge symmetry
of the string is $(0,1)$ superconformal invariance.  
Let us focus immediately on the simplest
heterotic string theory that can live in $D$ dimensions, 
type HO$^{+/}$.
Not counting ghosts and superghosts, the degrees of freedom 
on the string worldsheet are $D$ embedding coordinates
$X\uu\m$, their right-moving fermionic superpartners
$\psi\uu\m$, and $D+22$ left-moving current algebra
fermions $\llt\uu a$.
Physical states of the string correspond to 
normalizable states that are primary under 
the left-moving Virasoro algebra and the right-moving
super-Virasoro algebra.  Additionally, physical states must
have weight $(1,\hh)$.

The discrete gauge group of the worldsheet is a single, overall
$\IZ\ll 2$ that flips the sign on all fermions (both $\llt\uu a$ and $\psi\uu \m$)
simultaneously.  In other words, the partition function
on the torus corresponds to the diagonal modular invariant.  There
is a single Ramond sector, in which the $\psi\uu\m$ and
$\llt\uu a$ are simultaneously periodic.  These states correspond
to spacetime fermions that are spinors under $SO(D-1,1)$ and
$SO(D + 22)$.  The ground-state energy in this sector is
${1\over{16}}(D + 22)$: since all such states are
heavy, we will focus only on NS states.

For consistency, the total central charge of the matter must
be $(26,15)$.  The central charge of $D$ free scalars
along with $D$ right-moving and $D+22$ left-moving fermions 
is equal to $(26,15) + \lrdd {3\over 2}(D - 10), {3\over 2} 
(D - 10) \rrdd$.  We can cancel the central charge excess by
adding a dilaton gradient $V\ll + = V\ll - = - {q/ {\sqrt 2}}$,
where $q\sqd = {1\over{4\apr}} (D - 10)$.  The central charge
is then critical, and the worldsheet CFT defines a
classical solution to string theory that comprises a heterotic
version of the timelike linear dilaton theory.
Conformal invariance is automatic, since the worldsheet
theory is free and massless, with
action
\bbb
{\cal L}\ll{\rm kin} = 
 {1\over{2\pi}} G\ll{\m\n}
\lsqq
{2\over{\apr}} 
 (\pp\ll + X\uu\m)
(\pp\ll - X\uu\n)
-  i \psi\uu\m (\pp\ll - \psi\uu\n) \rsqq 
-  {i\over{2\pi}}  \tilde{\l}\uu a (\pp
\ll + \tilde{\l}\uu a)\ .
\een{hetfreelag}

\subsection{Giving a vev to the tachyons $\ct\uu a$}
We can deform the timelike linear dilaton background
by letting the tachyon acquire a nonzero value 
obeying the equations of motion.  The tachyon
transforms in the fundamental of $SO(D + 22)$.  A nonzero
tachyon vev $\ct\uu a (X)$ couples to the worldsheet as 
a superpotential
\bbb
W \equiv \llt \uu a : \ct\uu a (X) :\ ,
\eee
where the component action comes from integrating the
superpotential over a single Grassmann direction $\th\ll +$:
\be
{\Delta \cl} &\equiv& - {1\over{2\pi}} \int d\th\ll + ~ W
\nn \\ && \nn \\
&=& - {1\over{2 \pi}}
\lrdd
F\uu a :\ct\uu a(X) :  - i \sqrt{\apr \over 2} \cc 
: \pp\ll{\m} \ct\uu a (X):
\llt\uu a  \psi\uu \m  \rrdd\ .
\ee
The object $F\uu a$ is an auxiliary field, and the
kinetic term for the free, massless theory contains
a quadratic term ${1\over{2\pi}} F\uu a F\uu a . $ 
Integrating out $F\uu a$ results in a potential of the form
\bbb
\Delta {\cal L} = - {1\over{8\pi}} 
\sum\ll a
:\ct\uu a(X) \sqd: \ ,
\eee
and a supersymmetry transformation 
$\{Q, \tilde{\l}\uu a\} = F\uu a = \hh :\ct\uu a:$.\footnote{Above,
we have written
the normal-ordered form $ : \ct\uu a (X)\sqd :$
of the potential, rather than the 
singular form $\lrdd :\ct\uu a (X)  : \rrdd\sqd$.  In the
example we study, the difference between the two is canceled by
a second-order term in conformal perturbation theory that
is generated when two insertions of the Yukawa coupling approach
each other.  In terms of Feynman diagrams, this is just the
supersymmetric cancellation of the vacuum energy.}

The operator $:\ct\uu a(X) :$ must be a superconformal
primary of weight $(\hh, \hh)$.  Therefore, the equation
of motion for a weak tachyon field is
\bbb
\pp\uu \m\pp\ll\m \ct\uu a - 2 V\uu\m \pp\ll\m \ct\uu a
+ {2\over\apr} \ct\uu a = 0\ .
\eee
Even for perturbations satisfying the linearized
equations of motion,
there will generally be singularities 
in the product of a tachyon insertion with itself.
Consequently, there will arise a nontrivial conformal perturbation theory that
must be solved order by order in the strength of $\ct^a$.
However, there are special choices of $\ct^a$ for which
there are no singularities in the OPE of a superpotential
insertion with itself, and the perturbation is 
exactly marginal.

The simplest such choice is for the gradient of the tachyon to lie in
a lightlike direction:
\bbb
\ct\uu a = \m\uu a \exp{\b X\uu +}\ ,
\eee
with $q\b = {{\sqrt{2}}\over \apr}$.  The resulting 
bosonic potential is positive-definite and increasing
exponentially toward the future:
\bbb
\Delta
{\cal L} = - {1\over{8\pi}} (\m\uu a)\sqd ~\exp{2\b X\uu +}
+ {i\over{2\pi}} \sqrt{\apr\over 2} 
\b \m\uu a \exp{\b X\uu +} ~\llt\uu a \psi\uu + \ .
\eee
This theory has a potential barrier that is like a
Liouville wall moving to the left at the speed of
light, similar to the solution of \cite{previous}.
This theory may have interesting dynamics,
and should be studied further. 

The next simplest case is the one in which we allow the tachyon
to vary in $n$ dimensions $X\ll 2,\ldots , X\ll{n + 1}$ 
transverse to the light-cone dimensions $X\uu\pm$.  
We assume that the tachyon has an isolated zero
$\ct\uu {1} = \cdots = \ct\uu n = 0$ at $X\ll 2 = \cdots = 
X\ll{n + 1} = 0$, and we take the limit in which
the spatial wavelengths $k\ll 2\uu{-1},
\cdots,k\ll {n+1}\uu{-1}$ of the tachyons go to infinity.
We can then approximate the tachyon field by
\bbb
\ct\uu a (X) = 
\sqrt{2\over\apr}\cc
\exp{\b X\uu +} \lsqq
\sum\ll{b = 1}\uu n
\hat{\m}\uu a{}\ll b~X\ll{b+1} + O\lrdd k X\uu 2 \rrdd
\rsqq\ ,
\eee
with
\bbb
q \b = {{\sqrt{2}}\over\apr}\ .
\eee

Taking the long-wavelength limit for the tachyons amounts to
dropping the $kX\sqd$ terms.  The assumption of an
isolated zero is equivalent to assuming $\hat{\m}\uu a{}\ll b$
is nondegenerate.  For simplicity, let $\hat{\m} \uu a{}\ll b
\equiv \m ~ \d\uu a\ll b$.  The
superpotential is then 
\bbb
W = +\m\cc \sqrt{2\over\apr} ~ \exp{\b X\uu +}\sum\ll{a = 1}\uu n \llt\uu a
X\ll{a+1}\ ,
\eee
and the interaction Lagrangian appears as
\be
{\cal L}\ll{\rm int} &=& - {{\m\sqd}\over{4\pi}} \cc\exp{2\b X\uu +}\cc 
 \sum\ll{a = 1}\uu n : X\ll {a + 1}\sqd : 
\nn\\ && \nn\\
&& 
+ {{i \m}\over{2\pi}} \exp{\b X\uu +} \cc
\sum\ll{a = 1}\uu n \tilde{\l}\uu a
\lrdd  \psi\uu{a + 1} + \b X\ll{a + 1} \psi\uu + \rrdd\ .
\ee
This superpotential is the same as the one in \cite{hellerman1,
hellerman2}, except that the tachyon is increasing in a 
lightlike direction rather than a timelike direction.

The equations of motion are
\be
\pp\ll + \pp\ll - X\uu + &=& \pp\ll - \psi\uu + = 0 \ ,
\nn\\ && \nn\\
\pp\ll + \pp\ll - X\uu i &=& - {{\m\sqd}\over 4} 
\cc\exp{2\b X\uu +}\cc X\uu i \ , \qquad  i = 2,\cdots,n + 1\ ,
\nn\\ && \nn\\
\pp\ll + \pp\ll - X\uu - &=& + {{\b}\over 4} 
\cc\exp{2\b X\uu +}\cc
\biggl( \sum\ll{i = 2}\uu {n + 1} X\ll i X\ll i
\biggr)  
\nn\\ && \nn\\
&& - {{i \b\apr }\over 4}  \cc\exp{\b X\uu +}\cc
\sum\ll{a = 1}\uu{n }\tilde{\l}\uu a
\lrdd \psi\uu{a + 1} + \b \cc
X\ll{a + 1} \psi\uu + \rrdd\ ,
\nn\\ && \nn\\
\pp\ll + \tilde{\l}\uu a &=&
{{\m}\over 2} \cc\exp{\b X\uu +}\cc
\lrdd \psi\uu {a + 1} + \b X\ll{a + 1}\psi\uu + \rrdd \ ,
\nn\\ && \nn\\
\pp\ll - \psi\uu i &=& 
- {\m\over 2} \cc\exp{\b X\uu +}\cc \tilde{\l}\uu{i - 1}\ ,
 \qquad i = 2,\cdots,n + 1 \ ,
\nn\\ && \nn\\
\pp\ll - \psi\uu - &=& {{\b \m}\over 2} \exp{\b X\uu +} \cc\sum\ll{a = 1}\uu n
\tilde{\l}\uu a X\ll{a + 1}\ .
\eee

In infinite volume, the construction of the general classical
solution proceeds in parallel with
the case of the bosonic dimension-changing transition.
If we are interested in understanding the classical
behavior, we can set the fermions to zero.
Defining $f\ll\pm ,~{\bf F}\ll\pm$ as in Eqns.~\rr{littlefdef} and \rr{bigfdef},
we write
\bbb
X\uu + = \hh f\ll + (\s\uu +) + \hh f\ll - (\s\uu -)\ .
\eee
The general solution for $X\ll i$ is then
\bbb
X\ll i = \sum\ll \o A\ll {\o,(i)}\cc  \expr {{i\over 2} \o {\bf F}\ll + (\s\uu +)
+ {i\over 2}  {{\m\sqd}\over \o}  {\bf F}\ll - (\s\uu -)}\ .
\eee
Choosing a set of mode amplitudes $ A\ll {\o,(i)}$, one
can solve the
equations of motion for $X\uu -$ by treating it as Poisson's equation,
with  a fixed source proportional
to $\sum_i X\ll i\sqd \cc \exp{2 \b X\uu +}$.

To understand the classical
behavior of strings at late times,
we make the ansatz $\pp\ll{\s\uu 1} X\uu + = 0,$
and define $\numosc \ll {n(i)} (\s\uu 0)$ as the occupation numbers
for the modes of the $X\ll i$ fields.  Choosing $p\uu + > 0$, we have   
\bbb
X\uu + = \apr p\uu + (\s\uu 0 - \s\uu 0\ll{(0)}) + X\uu +\ll{(0)}\ ,
\xxx
\numosc\ll{n(i)}(\s\uu 0) \sim \numosc\ll{n(i)}\uu{\rm(fin.)}\ ,
\eee
for large values of $X\uu +$.  If we further assume that the 
$X\ll i$ are independent of $\s\uu 1$, then $\numosc\ll{i(n)} = 0$ for $n\neq 0$.  In this
case the behavior of $X\uu -$ at late times approaches
\bbb
X\uu - \sim 
\sum\ll i {{\numosc\ll{0(i)} \cc M(\s\uu 0)}\over{\b\apr p\uu{+2} \rws}}\ .
\eee

Any nonzero values for $N\ll{n(i)}$ ($n\neq 0$) 
can only contribute positively to the growth of $X\uu -$, 
by an amount increasing exponentially in time like $M$.
A particle therefore accelerates to the left, 
quickly approaching the speed of light,
unless all its $X\ll i$ oscillators are in their ground states.


The theory behaves in a simple fashion at the quantum
level as well. Just as for the dimension-changing transition in the bosonic string,
the renormalization of the dilaton and metric
occur only at one-loop order on the worldsheet, where only
the massive fields
$\llt \uu a,~ X\ll{a+1},~ \psi\uu {a+1}$
circulate in the loop.  

Integrating out a massive Majorana fermion of mass 
$M (X)$ gives the shifts
\bbb
\Delta \Phi = + {1\over{12}} \ln \lrdd {M\over{\tilde{\m}}} 
\rrdd\ ,
\qquad 
\Delta G\ll{\m\n} = + {{\apr}\over{12}} {{\pp\ll\m M  \pp\ll\n M}
\over{M\sqd}}\ ,
\eee
at one-loop order.
Here, $M(X)$ is allowed to depend arbitrarily on the
target space coordinates
$X\uu\m\in\{X\uu\pm,X\ll{n+2},\cdots,X\ll{D-1}\}$,
which are not being integrated out.
Using the formulae in Eqns.~\rr{dilrenormone} and
\rr{metrenormone},
we see that the contribution from the massive boson
is twice that for the massive fermion.
The total contribution from all the massive $X,~\psi$ and $\llt$
fields is therefore
\bbb
\Delta \Phi = + {n\over{4 }} \ln \lrdd {M\over{\tilde{\m}}} \rrdd\ ,
\qquad
\Delta G\ll{\m\n} = + {n\over{ 4}} {{\pp\ll\m M  \pp\ll\n M}
\over{M\sqd}}\ .
\eee
Letting
\bbb
M \equiv \m\cc\exp{\b X\uu +}\ ,
\een{defofm}
we obtain
\bbb
\Delta V\ll + = + {{n\b}\over{4 }}\ ,
\qquad
\Delta G\ll{++} = + {{n\b\sqd\apr}\over{ 4}}\ , 
\qquad
\Delta G\uu{--} = - {{n\b\sqd\apr}\over{ 4}}\ ,
\eee
with all other components of $G\ll{\m\n}$ and $V\ll\m$ unrenormalized.

One difference from the dimension-changing domain wall
in the bosonic string is that the potential energy
need not be fine-tuned by hand.  Since we are integrating
out fields in complete multiplets, the quantum
correction to the potential energy vanishes.
Even nonperturbatively, the effective
potential at $X\ll 2 = \cdots = X\ll {n + 1} = 0$ must
vanish.  Any correction to the effective potential
must come from an effective superpotential term containing
a single $\llt$ fermion.  The remaining fermions are
$\llt\uu{n + 1},\cdots ,\llt\uu{D + 22}$.  Any effective
superpotential with a single $\llt$ fermion would violate
the residual $SO(D + 22 - n)$ symmetry left unbroken by
the tachyon vev.

The linear dilaton central charge at $X\uu + \to + \infty$ is
\bbb
c\uu{\rm dilaton} = 6\apr \hat{G}\uu{\m\n}
\hat{V}\ll\m \hat{V}\ll\n 
= - 6\apr q\sqd + {{3 n q\b\apr}\over{\sqrt{2}}}
- {{3 n \b\sqd q\sqd \apr\sqd}\over 4}\ .
\eee
Using
$q\sqd = {1\over{4\apr}} (D - 10)$ and $q\b = \sqrt{2} / \apr$,
we find that the final dilaton central charge is
\bbb
c\uu{\rm dilaton} = {3\over 2} \lsqq - (D - 10) + n \rsqq\ .
\een{finaldilhet}
Thus, exactly ${{3n}/ 2}$ units
of central charge are transferred from the $n$
massless multiplets $X\ll{a + 1},~ \psi\uu{a + 1}$ and
current algebra fermions $\llt\uu a$ 
to the dilaton gradient at $X\uu + = + \infty$.
The total central charge is again the same at $X\uu + 
= + \infty$ as at $X\uu + = - \infty$, and in particular
equal to $(26,15)$.

\subsection{Transitions among nonsupersymmetric heterotic theories}
\begin{figure}[htb]
\begin{center}
\includegraphics[width=3.2in,height=3.2in,angle=0]{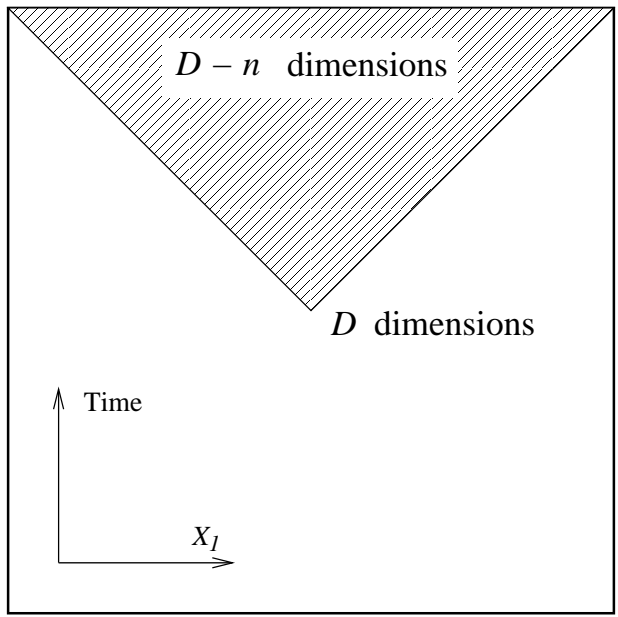}
\caption{The dynamical transition from $D$ dimensions outside 
a bubble wall to $D-n$ dimensions in the interior.  Our solution
focuses on the upper left-hand corner of the diagram, where the
bubble is a domain wall moving to the left at the speed of light.}
\label{bubble}
\end{center}
\end{figure}

In the case where the spatial slices 
are simply $\IR\uu {D-1}$, we can choose $n$ to be 
any integer from $1$ to $D-2$.  We always end up 
at $X\uu + = + \infty$ 
with a consistent string theory in $D - n$ dimensions with
total central charge equal to $(26,15)$.
If $n < D - 10$, the final theory is another
supercritical theory.  If $n > D- 10$, we end up with
a subcritical theory with a spacelike linear dilaton.
In both cases there is a tachyon in the fundamental
representation of the unbroken gauge group $SO(D + 22 - n)$.
If $n = D - 10$, our final theory is a critical, unstable
heterotic string theory with gauge group $SO(32)$ and a
lightlike linear dilaton rolling to weak coupling in the
future.   The spacetime transition from $D$ to $D-n$ 
dimensions is depicted schematically in Fig.~\ref{bubble}.

\ \\

We can also choose the maximal number
of transverse dimensions $n = D - 2$.
In this case, the tachyon in the final theory is stabilized
due to its modified dispersion relation in the background
of the linear dilaton \cite{seiberg}.
This theory has unbroken gauge
symmetry $SO(24)$, and a finite number of local degrees of
freedom (namely, a set of scalar tachyons in the fundamental
representation of $SO(24)$).

In other words, 
the tachyon in the $2D$ string theory
is precisely zero.
 The end of the throat in
the $2D$ theory is at infinitely strong coupling,
with no Liouville wall to bar string states from the
strong coupling region; this theory was studied
in \cite{newhat}.  At a generic point in the 
moduli space of the $2D$ theory, the tachyon
condensate is nonzero, and the maximum coupling
attainable by a string is finite.  This
perturbation cuts off the strong
coupling end of the theory with a Liouville
wall, and is a valid perturbation of
the pure $2D$ string theory.  However,
we do not know how this perturbation may lift to a deformation
of our full solution.

The $2D$ heterotic background was analyzed recently in
\cite{seiberghet1, seiberghet2} and found to have a nontrivial
phase structure, as well as a characteristic nonlocal behavior not
familiar from field theory or other string theories.  It would
be interesting to understand the ways in which the 
peculiar properties of the $2D$ heterotic theory might descend
from the properties of the higher dimensional theory that
from which it decays via closed string tachyon condensation.

\subsection{Return to the supersymmetric vacuum in $D = 10$ with gauge group $SO(32)$}
Another interesting option is to orbifold the spatial slices
$\IR\uu{D - 1}$
by a $\IZ\ll 2$ involution that acts on $D-10$ of the
spatial coordinates $X\ll 2, \ldots , X\ll{D - 9}$,
leaving only $9$ spatial dimensions and the time dimension
unorbifolded. 
%
%
Modular invariance constrains the action of the 
involution on current algebra fermions $\llt\uu a$.  The simplest
modular invariant choice is to act with a $-1$ on all
$D + 22$ of the $\llt\uu a$.  This singularity
was studied in \cite{hellerman1} and referred to as an
orbifold singularity of \it stable type. \rm

This choice preserves the full $SO(D + 22)$ gauge symmetry,
and leaves massless fermions propagating on the $9+1$-dimensional
fixed locus of the $\IZ\ll 2$ reflection.  The localized
massless spectrum includes a spacetime fermion
transforming as a massless spin-${3\over 2}$ field under the
9+1 dimensional longitudinal Poincare group.  The orbifold
projection imposes a boundary condition forcing the tachyons
$\ct\uu a$ to vanish at the fixed locus.  

Upon perturbing the system with a tachyon vev
\bbb
\ct\uu a (X) =\sqrt{2\over\apr} \cc \hat{\m}\uu a{}\ll b ~\exp{\b X\uu +}~
X\ll{b + 1}\ ,
\eee
one needs the superpotential to be even under the orbifold
action.  
We may therefore use only odd
coordinates $X_{b+1},~ b\in\{1 ,\cdots,D - 10 \}$.  The
mass matrix $\hat{\m}$ thus has rank $n$ at most
$D - 10$, and we will take it to be $\m \, \mathbbm{1}\ll{\rm n\times n} .$

The maximal value $n = D-10$ is particularly interesting.
In this case, the final theory is critical, with lightlike linear dilaton
and continuous gauge group $SO(32)$.
The GSO projection inherited by the critical theory is
$\IZ\ll 2 \times \IZ\ll 2$, generated by
worldsheet fermion
number mod two,
$g\ll {\rm F} \equiv (-1)\uu{F_{W}}$ and
$g\ll {\rm L}\equiv (-1)\uu{\# \llt}$, which is the center of
the unbroken $SO(32)$.  The bosons and right-moving fermions
on which $g\ll {\rm L}$ acts nontrivially are massive, and 
decoupled at $X\uu +\to + \infty$.  Likewise, after we
integrate out massive fields at $X\uu +\to + \infty$,
$g\ll{\rm R} \equiv g\ll{\rm F} g\ll{\rm L}$ acts only
on right-moving fermions $\psi\uu\m$.  The $\IZ\ll 2 \times
\IZ\ll 2$ GSO projection is precisely the {\it
spacetime-supersymmetric } GSO projection of the 
$SO(32)$ heterotic string.  The gravitini that generate
spacetime supersymmetry transformations are the 
massless ground states of the sector 
twisted by $g\ll{\rm R}$.  

The final state of our theory is therefore a state of the
supersymmetric heterotic string with $SO(32)$ gauge group.
Indeed, the final state in this example is actually a BPS solution
in the limit $X\uu + \to \infty$.  To see this, recall
the supersymmetry transformations for the heterotic string 
\cite{chs}
\be
\d \Psi\ll\m &=& \lsqq \gg\ll\m + {1\over 4}
\omega\ll\m \uu A{}\ll B
\G\ll A{}\uu B
 - {1\over 4}
H\ll{\m\n\s} \G\uu{\n\s} \rsqq \e \ , 
\nn\\ && \nn\\
\d \L &=&  \lsqq \G\uu \m (\gg\ll\m \Phi)
- {1\over 6} H\ll{\m\n\s}\G\uu{\m\n\s} \rsqq \e\ ,
\nn\\ && \nn\\
\d \chi\uu{[ab]} &=& F\ll{\m\n}\uu{[ab]} \G\uu{\m\n} \e\ .
\ee
For flat string-frame metric, vanishing $H$-flux and
trivial $SO(32)$ gauge field, the supersymmetry transformations
reduce to
\bbb
\d \Psi\ll\m = \pp\ll \m \e \ ,
\qquad
\d \L =  \G\uu \m (\pp\ll\m \Phi) \e \ ,
\qquad
\d \chi\uu{[ab]} = 0\ .
\eee
For a lightlike linear dilaton, the matrix $\G\uu\m
(\pp\ll\m \Phi)$ is nilpotent of order $2$, and
annihilates one half of the spinors on which it acts.
The kernel can be taken to be constant spinors, so
the gravitino variation vanishes.
The gaugino variations vanish trivially, since the 
gauge field strength is turned off.

The late-time behavior of the solution is that of a state
preserving 8 of 16 supersymmetries.  
We have found an exact solution connecting supercritical 
heterotic string theory in $D > 10$ dimensions with a
half-BPS state of the critical, 10-dimensional $SO(32)$
heterotic string theory.

\section{Dimension-changing transitions in type 0 string theory}
We can also study transitions from supercritical type 0 string theory
to type 0 or type II string theory in lower dimensions.
As in the heterotic case, starting with $\IR^{D-1}$ spatial slices gives
rise to an unstable theory in lower dimensions.  By starting with
orbifolded spatial slices, it is possible to reach a final state 
that is spacetime supersymmetric.

The local worldsheet gauge symmetry in type 0 string theory
is $(1,1)$ superconformal invariance.  
Not counting ghosts and superghosts, the $2D$ fields
on the string worldsheet are $D$ embedding coordinates
$X\uu\m$ and their fermionic superpartners $\pst\uu\m,~ \psi\uu\m$,
and the physical states of the string are normalizable,
super-Virasoro primaries of weight $(\hh,\hh)$.
The total central charge of the matter must
be $(15,15)$, and the central charge contributed 
by $D$ free massless fields $X\uu\m$ and their superpartners
is $(15,15) + \lrdd {3\over 2}(D - 10), {3\over 2}
(D - 10) \rrdd$.
The central charge excess can be canceled by
adding a dilaton gradient $V\ll + = V\ll - = - {q/ {\sqrt 2}}$,
where
\be
q\sqd = {1\over{4\apr}} (D - 10)\ .
\ee 
The worldsheet CFT then defines a supercritical type 0 string theory,
with worldsheet action
\bbb
{\cal L}\ll{\rm kin} = 
 {1\over{2\pi}} G\ll{\m\n}
\lsqq
{2\over{\apr}} 
 (\pp\ll + X\uu\m)
(\pp\ll - X\uu\n)
-  i \psi\uu\m (\pp\ll - \psi\uu\n) -  i \pst\uu\m (\pp
\ll + \pst\uu\n) \rsqq\ .
\een{type0freelag}

\subsection{Description of type 0 string theory in $D > 10$}
The discrete gauge group of the worldsheet is a single, overall
$\IZ\ll 2$ acting with a $-1$ simultaneously on all fermions
$\psi,~\pst$.  This is just the 
symmetry $\mfw$ of worldsheet fermion number mod 2, which must
be gauged in any consistent theory involving worldsheet
fermions.  There is also an apparent global symmetry acting 
on $\pst$ with a $-1$ and on $\psi$ with a $+1$.
We call this symmetry `apparent' because it is
violated in type 0 string theory in odd dimensions,
as will be demonstrated below.
This symmetry, which we refer to as $\mflw$ (the `L' indicates ``left''),
has no analog in the physical spectrum
of type II string theories because it is gauged; it becomes
part of the GSO group, along with $\mfw$. 

All type 0 theories have a
single real tachyon $\ct$, corresponding to the
oscillator ground state of the NS/NS sector.  
The tachyon $\ct$ couples to the string worldsheet as
a superpotential, so it must be odd under
$\mflw$ if $\mflw$ is indeed a symmetry.
In addition to the NS/NS sector,
there is a R/R sector in which the 
worldsheet fermions are 
all periodic, and states correspond
to $p$-form fields.  In even dimensions, the 
massless content of the R/R sector is either $[0]\sqd \oplus
[2]\sqd \oplus \cdots \oplus [{D\over 2} - 3 ]\sqd
\oplus  [{D\over 2} - 1 ]\uu 1$ or
$[1]\sqd \oplus
[3]\sqd \oplus \cdots \oplus [{D\over 2} - 2 ]\sqd$,
depending on the sign choice of the GSO projection.
Here $[p]$ denotes a massless $p$-form potential, and
the superscript denotes the multiplicity.
In odd dimensions there is only one choice of GSO projection,
and the spectrum of $p$-form fields is
$[0]\uu 1 \oplus [1]\uu 1 \oplus [2]\uu 1 \oplus \cdots \oplus
[{{D - 3}\over 2}]\uu 1 $.

In odd dimensions the symmetry $\mflw$ is anomalous as a
global symmetry on the string worldsheet.  There is no consistent assignment of
transformations to spin fields for which the OPE preserves
any chiral R-parity.
To see this, notice that the R/R $p$-forms have vertex operators that
are spin fields for an odd number of left-moving
and an odd number of right-moving worldsheet fermions.
Likewise, the OPE of two such
spin fields closes on NS/NS fields with an odd number 
of left- and an odd number of right-moving worldsheet
fermions.\footnote{This becomes apparent if one considers the
fermionic sector of the CFT as a product of $D$ critical Ising models.}
There are therefore nonvanishing three-point
functions that would violate the $\IZ\ll 2$
symmetry for any choice of action on the R/R form.
An example would be the three-point function
involving two R/R axions and a tachyon.

Since the tachyon is odd under $\mflw$,
the symmetry $\mflw$ cannot extend as a $\IZ\ll 2$ symmetry to
act in a consistent way on the R/R scalar as well.  One could
try to extend $\mflw$ to generate a $\IZ\ll 4$ symmetry,
but this would mean it would have to act on the R/R zero-form
with eigenvalue $\pm i$.  The complex conjugate of the R/R axion
would then have to have the conjugate eigenvalue.  Since the
R/R axion is a real field, however, this is not possible.

Another way to see the absence of a conserved $\mflw$ is to
consider amplitudes on the torus, with the odd spin structure
$(P,P)$ for the worldsheet fermions.  In $D$ dimensions, this spin 
structure has $D$ zero modes for the right-moving fermions and
$D$ zero modes for the left-moving fermions.  There are therefore
nonvanishing amplitudes with $D$ tachyon vertex operators inserted.
If $D$ is odd, there is no way for the tachyon to be odd under 
a conserved $\mflw$.

Since we gain our control over tachyon condensation 
in the type 0 theory by using the exact global symmetry $\mflw$, we will
focus on even dimensions $D = 10 + 2 K$. 
Though it is not in general possible to {\it orbifold  } by 
$\mflw$ (i.e.,~to treat it as part of the GSO group),
it is always a good global symmetry of the worldsheet in
even dimensions, and therefore an exact discrete gauge symmetry
in spacetime.  Any superpotential $W$ preserving $\mflw$ must
be odd under it, since the superspace measure $\int d\th\ll +
d\th\ll -$ is also odd.  $\mflw$ is therefore a discrete
$R$-parity, and its existence will allow us to constrain tachyon
condensates in a natural way, without fine-tuning.

\subsection{Giving a vev to the tachyon $\ct$}
We wish to deform the background by a tachyon expectation value
that obeys the equations of motion.  The tachyon couples
to the worldsheet as a superpotential, integrated over
one Grassmann variable $\th\ll \pm$ of each chirality:
\be
\Delta {\cal L} &=& {i\over{2\pi}}  \int d\th\ll + d\th\ll - \ct(X)
\nn \\ && \nn \\
&=& - {1\over{2\pi}} \cc 
\sqrt{\apr\over 2}\cc F\uu\m : \pp\ll\m \ct (X): 
+ {{i\apr}\over{4\pi}} :\pp\ll\m \pp\ll\n \ct (X) : \pst\uu\m
\psi\uu\n \ .
\ee
Including the
${1\over{2\pi}} F\uu\m F\ll\m$ term that comes along with
the kinetic term, and integrating out the auxiliary field $F\uu\m$,
we find the resulting potential
\bbb
- \Delta {\cal L} = {\apr\over{16\pi}}
G\uu{\m\n} (:\pp\ll{\m}\ct (X) : ) (:\pp\ll\n \ct(X):)\ .
\eee
The condition for conformal invariance of the linearized
perturbation is that $:\ct(X):$ be primary of
weight $(\hh,\hh)$.  This amounts to the condition
\bbb
\pp\sqd \ct - 2 (V\cdot \pp) \ct + {2\over\apr} \ct = 0\ .
\eee

Our exactly solvable examples have $\ct$ appearing with exponential
dependence on a lightlike direction.
The {\it simplest} case to consider is the case where $\ct = \exp{\b
X\uu +}$ and $\b q = \sqrt{2} / \apr$.  This perturbation
is conformally invariant and gives rise to an
exactly solvable background describing a
transition between two different string
theories.  It does not reduce the
number of spacetime dimensions, but instead changes the {\it kind
of string theory altogether.} We describe this solution in detail
in a separate paper \cite{paper3}.

Searching for exactly solvable models of 
dynamical dimension change, it turns out that the
easiest models to control are those for which
the vev of the tachyon preserves some spacetime reflection symmetry,
combined with the chiral $R$-parity $\mflw$.  We
thus look for perturbations such that the superpotential
$W$ is odd under the reflection of some subset
of the coordinates $Y\uu{b}\equiv X\ll{b + 1},$
with $b = 1,\ldots ,~n$:
\bbb
\ct = \sum \ll{b = 1} \uu n :{{\tilde{f}\ll b (X)}\over{k\ll b}} \cc\sin{(k\ll b Y\uu b)} :\ ,
\eee
were the $\tilde{f}\ll b$ depend only on
$X\ll 0,~X\ll 1 ,~ X\ll{n + 2} , \ldots,~
X\ll{D - 1}$.
Taking $k\ll a\to 0$, we obtain the limiting configuration
\bbb
\ct = \sum \ll{b = 1} \uu n :\tilde{f}\ll b (X) Y\uu b :\ .
\eee
We can further restrict the form of $\tilde{f}\ll b$ by
adopting the following ansatz: 
\bbb
\tilde{f}\ll b \equiv \exp{\b\ll b  X\uu +} \cc : f\ll b (X) :\ ,
\eee
where $f\ll b(X)$ depends only on $X\ll {n + 2},\ldots,~X\ll {D - 1}$.

Treating the worldsheet dynamics semiclassically, the potential takes the form
\bbb
 - 2\pi\cc \Delta {\cal L} &=& 
\sum\ll{b = 1}\uu n \exp{2\b\ll b X\uu +}
f\ll b(X)\sqd 
\nn\\
&&
\nn\\
&& + \sum\ll{a,b,c = 1}\uu n 
\exp{(\b\ll a + \b\ll b) X\uu +}
Y\uu a Y\uu b f\ll{a,c}(X) f\ll{b,c}(X) \ .
\een{semiclassicalpotential}
Let us analyze the classical supersymmetric moduli space
of the worldsheet theory.
Worldsheet supersymmetry is unbroken at
exactly those values $X,~ Y$ for which $\{f\ll b (X) = f\ll{b,c}(X) Y\uu c = 0\}$.
This contains the locus ${\bf M} \equiv \{f\ll b (X) = Y\uu c = 0\}$.  
If ${\bf M}$ is a smooth manifold, then there are no other components.  
Some linear combinations of $Y\uu c$ can have classical flat directions over
singular points in ${\bf M}$.

For a generic choice of functions $f\ll b(X)$, the space
${\bf M}$ will indeed be smooth.
So the 
classical supersymmetric vacuum manifold at fixed $X\uu{0,1}$
defines a smooth submanifold of real
codimension $n$ in the $\IR\uu {D - 2 -n}$-space
spanned by $X\ll{n + 2} ,\ldots,~ X\ll{D - 1}$.  Treating
the dynamics semiclassically, the theory at $X\uu + \to + \infty$
is a sigma model on ${\bf M}$, times $\IR\uu{1,1}$.
To the extent that our semiclassical treatment signals the
existence of a well-defined CFT, 
the full $2$D field theory, with $X\uu\pm$ included,
describes a transition from type 0 string theory on
$\IR\uu{D-1,1}$ to type 0 string theory on $\IR\uu{1,1} \times {\bf M}$, where
the dimensionality of ${\bf M}$ is $D - 2 - 2n $. 

Such a CFT 
describes a transition in which
the string theory begins in a primordial, cosmological state
at early times and chooses a topology dynamically,
through closed string tachyon 
condensation.\footnote{A version of this general construction
exists also for the unstable heterotic string
discussed in the previous section.  In that case,
we can choose $\ct\uu b(X) = f\ll b(X)$, with $b = 
1,\ldots, n$.  Treated semiclassically
in $\apr$, the endpoint is a heterotic string theory
with gauge group $SO(D+22 - n)$ propagating
on $\IR\uu{1,1} \times 
\left( {\bf M} = {{\bigcap} \atop {b = 1}} \{f\ll b = 0\}\right)$.
The gauge bundle in the lower-dimensional string
theory is simply the tangent bundle of ${\bf M}$
times a trivial bundle.}

In general, worldsheet quantum effects on the worldsheet will alter this picture substantially.
The on-shell condition for the linearized tachyon perturbation is
\bbb
\pp\sqd \ct - 2 V\cdot \pp \ct + {2\over\apr} \ct = 0\ ,
\eee
which is satisfied if and only if
\bbb
\pp\ll i \sqd f\ll b(X) = - k\ll b \sqd f\ll b(X) \equiv 
\lrdd - {2\over\apr} + \sqrt{2} \b\ll b q \rrdd  f\ll b(X)\ .
\eee
By superposing different $f\ll b$,
tuning coefficients and taking the long-wavelength limit $k\ll b\sqd \apr \to 0$,
one can obtain superpotentials of the form $\exp{\b X\uu +}~ Y\uu b ~
{\bf P}\ll b (X\ll i, X\uu +)$,
where ${\bf P}\ll b (X\ll i, X\uu +)$ are polynomials in the $X\ll i$ and $X\uu +$.
At fixed $X\uu +$, the locus ${\bf M}$ is an arbitrary real algebraic variety.  The
$X\uu +$ encode the leading logarithmic corrections to the size and shape of 
the manifold ${\bf M}$ under renormalization group flow.  The $X\uu +$ dependence
of the shape can vanish only if $\pp\ll i \sqd {\bf P}\ll b  = 0$.

We consider the simplest case where the above condition holds, namely
when the $f\ll b(X)$ are 
linear functions of $X$:
\bbb
f\ll b (X) = \sum\ll{a = 1}\uu{n }
 \hat{\m}\uu a{}\ll b X\ll {n + 1 + a} \ .
\eee
The condition that ${\bf M}$ be
nonsingular reduces to the condition
that
$\hat{\m}\uu a{}\ll b$ have nonvanishing
determinant.  For simplicity, we take
$\hat{\m}$ equal to $\m\cc{\mathbbm 1}\ll
{{\rm n}\times {\rm n}}$:
\bbb
\hat{\m}\uu a{}\ll b \equiv \m\cc \d\uu  a{}\ll b \ ,
\eee
in which case the full superpotential is
\bbb
W = \ct = \m\cc\exp{\b X\uu +} 
\sum\ll{b = 1} \uu n  \cc Y\uu b X\ll{n + 1 + b}\ ,
\qquad
q \b = {{\sqrt{2}}\over\apr}\ .
\eee

This CFT is exactly solvable,  both at the classical
and at the quantum level.  The analysis 
of the general classical solution
on a noncompact worldsheet proceeds
completely in parallel with the analysis of the
dimension-changing transition in the bosonic
string and the heterotic string.  The fields $X\uu +,~
\psi\uu +,$ and $\pst\uu +$ are free, and the massive
fields $Y\uu b$ and $X\ll{n + 1 + b}$ (and their superpartners) 
obey free equations of motion with time-dependent mass
terms, proportional to exponentials
of $X\uu +$.

The study of the classical equations of motion
works the same in this example as in the bosonic and
heterotic examples.  As in those cases, the classical potential
is simply a number of scalar fields $X\ll {b + n + 1},~Y\uu b,$ 
with masses equal to $\m\cc\exp{\b X\uu +}$.
Taking the ansatz $\pp\ll {\s\uu 1} X\uu + = 0$,
we can solve for
the trajectories of the string in terms of Bessel functions, and
we find exactly the same behavior as in the bosonic
and heterotic cases: states of the string with massive fields
excited have their energy locked permanently into
the massive modes by virtue of the adiabatic theorem.  They
are then pushed along the bubble wall for the rest of
time.  States that have no massive modes
excited can freely penetrate into the interior of the
tachyonic phase, where they move in $2n$ fewer dimensions.

The simplifications at the quantum level that hold for
the bosonic and heterotic worldsheet theories, hold in the CFT
of this section as well.  All connected tree and loop graphs consist of
a single open or closed massive line, with lightlike
lines emanating from some number of vertices placed along the line.
In particular, the couplings that
define the metric and dilaton receive corrections only from
one-loop graphs, when the massive degrees of freedom are integrated out.
The difference here is that the massive multiplets running in the loop are
larger, containing the fields $X\ll {n + 1 + b},~
Y\uu b,~ \psi\uu {X\ll{n + 1 + b}} ,~
\psi\uu {Y\uu b},~   \pst\uu {X\ll{n + 1 + b}}$ and
$\pst\uu {Y\uu b}$ for all $b = 1,\ldots,n$.
The outgoing lines attached to the massive loop
are all ``$+$'' lines, just as in the heterotic
and bosonic examples.  In the bosonic case there
were only outgoing $X\uu +$ lines, while outgoing $\psi\uu +$ 
lines were also allowed in the heterotic case.  In the type $0$ examples,
outgoing lines attached to a loop include the fields
$X\uu +,~ \psi\uu +$ and $\pst\uu +$.

In the case where the spatial slices 
are simply $\IR\uu {D-1}$, we can choose $n$ to be 
any integer in the range $n\in 1,\ldots ,~{D\over 2}-1$.  At $X\uu + = + \infty$
we arrive at a consistent string theory in $D - 2 n$ dimensions, with
total central charge equal to $15$.
If $2 n < D - 10$, the final theory is another
supercritical theory, while for $2 n > D- 10 $ we end up with
a subcritical theory with a spacelike linear dilaton.
If $2 n = D - 10$, the final theory is a critical, unstable
string theory of type $0$ with a
lightlike linear dilaton rolling to weak coupling in the
future. 
This theory has 
an unbroken $\IZ\ll 2$ symmetry under which
the tachyon is odd. The tachyon in the $2D$ string theory
is identically zero.  The $2D$ type 0 theory can be deformed
by the addition of a Liouville wall in the usual way \cite{newhat}.
As for the $2D$ heterotic case, we do not know what deformation of
our full, time-dependent solution would produce a
Liouville wall in the final state.

\subsection{Return to the supersymmetric type II vacuum in $D = 10$}
We can also consider orbifolding the spatial slices in 
a particularly interesting way.   As in the heterotic case, 
allowed orbifold singularities are
highly constrained by the requirement of
modular invariance, which manifests itself
as the need for level matching in twisted sectors.
For the type 0 string in $D\equiv 10 + 2K$ dimensions, the simplest
consistent orbifold action is a $\IZ\ll 2$
generated by an operation $g\ll {\rm L}$.  We define
$g\ll{\rm L}$ to 
act by reflection of the 
$K$ real coordinates $Y\uu b$, as well as 
acting by the
symmetry $\mflw$, which
acts as $-1$ on the left-moving worldsheet
supercurrent.
In other words,
in addition to its geometric action, $g\ll {\rm L}$ is a chiral
$R$-parity.  The full GSO/orbifold group is then
$\{1,~\mfw,~ g\ll{\rm L},~ g\ll{\rm R}\equiv \mfw \cdot g\ll{\rm L}\}$.

The resulting singularity $Y\uu b = 0$ preserves a Poincare invariance of the 
$\IR\uu{9+K,1}$ longitudinal to the fixed locus,
parametrized by $\{X\uu 0,~ X\ll 1,~ X\ll{K + 2},
\ldots,~ X\ll{2K+9}\}$.  This symmetry
is broken only by the
timelike linear dilaton.  There is also an $SO(K)$ rotational invariance
transverse to the fixed locus.  In the NS/NS and R/R sectors the coordinates
$Y\uu b$ are untwisted, while in the R/NS and NS/R sectors the coordinates $Y\uu b$
are twisted.  That is, spacetime fermions live only on the 
geometric fixed locus, and spacetime bosons propagate only in the
bulk.

Now we calculate the ground state weights in the
various sectors, ignoring the contributions of $\psi\uu{0,1},~\pst\uu{0,1}$
fermions.\footnote{We eliminate these as usual, using
the gauge freedom of local worldsheet supersymmetry.}
The oscillator ground states 
in the R/NS sector have weight $\left( {K\over 8} + \hh , {K\over 8}  \right)$.
The GSO and orbifold projections in the twisted R/NS sector
can be satisfied by acting with a single fermionic oscillator $\psi\uu{X\uu\m}$,
with $\m = 0,~1,~ K+2,\ldots,~2K+9$, which gives a level-matched state of weight
$\lrdd {K\over 8} + \hh , {K\over 8}  + \hh \rrdd$.  The ground states
in the twisted R/NS sector make up the states of a massless vector-spinor $\Psi\uu\m\ll{\a p}$,
where $\a$ is a spinor index of $SO(9+K,1)$ and $p$ is a spinor index of 
the transverse $SO(K)$.  The vector index $\m$ runs only along the
longitudinal directions $\m =  0,~1,~ K+2,\ldots,~2K+9   $. 
The NS/R sector 
is also level-matched, with the same content as the R/NS sector.

The GSO projection cuts the multiplicity of the localized fermions
by imposing the condition
\bbb
\lrdd \G\ll{\rm SO(9+K,1)} \otimes \g\ll{\rm SO(K)} \rrdd \cdot \Psi\uu\m 
	= \pm \Psi\uu\m\ ,
\eee
where $\G\ll{\rm SO(9+K,1)}$ and $\g\ll{\rm SO(K)}$ are products of
all gamma matrices of the longitudinal $SO(9+K,1)$ and the transverse $SO(K)$,
respectively.  The $\pm$ sign can be chosen independently
in the NS/R and R/NS sectors.  When $K$ is even, these are simply the usual
chirality matrices, up to a phase.  When $K$ is odd, there is
no chirality for spinors of $SO(9+K,1)$ or $SO(K)$, but there is
still a combined condition that
cuts the complex Dirac spinor
representation by two.  The precise meaning of this condition depends 
on the value of $K$, mod eight, and can be read off the table and discussion in
Appendices A.3 and A.4 of \cite{hellerman1} (with $n$ in that 
reference replaced by $K$ here.)
There is also a reality condition naturally derived from
the reality of the worldsheet fermions.  The details of
this condition also depend on the value of $K$ mod
eight (for further details, the reader is referred to Ref.~\cite{hellerman1}).

Apart from generating an interesting spectrum of massless fermions
propagating on the fixed locus, 
the other effect of the orbifolding is to impose boundary conditions
on the bulk fields at the orbifold singularity $Y\uu b = 0$.  In
particular, the superpotential $W$ must be odd under any
chiral $R$-parity.  If $W$ is a function of $X,~Y$ only, it must therefore be
odd under $Y\uu b \to - Y\uu  b$.

Since any allowed worldsheet superpotential must be 
odd under $g\ll{\rm L}$, the tachyon must be
odd under reflection of $Y$:
\bbb
\ct(X,-Y) = - \ct(X,Y)\ .
\eee
In particular, the tachyon must vanish at the origin
of the $Y$ coordinate.
The bosonic potential must therefore have a minimum at 
$Y = 0$ with zero energy, just as in the
heterotic case.  

By pairing $Y$ and $X$ coordinates with
mass terms in the superpotential, we can 
eliminate a maximum of $2K$ dimensions: 
all $K$ of the $Y$ dimensions can pair with
$K$ of the $X\uu\m$ directions and give them mass.  The
minimum dimension of the final state is therefore the critical dimension 
$D_{\rm crit} = 10$.  
By the same calculation
we performed in the heterotic case (see Eqn.~\rr{finaldilhet}),
the metric and dilaton change by
\bbb
\Delta V\ll + = + {{n\b}\over{2 }}\ ,
\qquad
\Delta G\ll{++} = + {{n\b\sqd\apr}\over{ 2}} \ ,
\qquad
\Delta G\uu{--} = - {{n\b\sqd\apr}\over{ 2}} \ .
\eee
The renormalized dilaton central charge is
\bbb
c\uu{\rm dilaton} = 3 (n - K)\ .
\eee
When $n = K$, the final state has $10$ dimensions,
the dilaton central charge vanishes
and the readjusted dilaton is null
with respect to the readjusted metric.
Moreover, the
GSO projection is that of type II string theory
rather than type 0.
Since all degrees of freedom in the
$Y$ multiplets have become
massive and decoupled, the orbifold symmetry
$g\ll{\rm L}$ acts only on the remaining
left-moving fermions $\pst\uu\m$.  That is,
$g\ll{\rm L}$ becomes equal to $\mflw$.   Therefore
the GSO group in the final
state at $X\uu +\to \infty$
is generated by $\mflw$ and $\mfw$, which means the
effective GSO projection and Ramond
sector content is exactly that of a type II string, rather
than type 0 string.
The final vacuum is the critical, supersymmetric
vacuum of type II string theory, with the choice
of IIA or IIB depending on 
whether the initial supercritical state is type 0A or 0B.

In the absence of flux and curvature, the supersymmetry 
transformations of the gravitini and dilatini in
type II string theory are
\bbb
\d \Psi\ll\m\uu{(1,2)} = \pp\ll \m \e\uu{(1,2)} \ ,
\qquad
\d \L\uu{(1,2)} =  \G\uu \m (\pp\ll\m \Phi) \e\uu{(1,2)} \ .
\eee
For a final dilaton gradient $\pp\ll\m \Phi$ that is
constant and lightlike,
there are exactly 16 supercharges preserved, out of
the full 32.  Our solution therefore describes the
dynamical relaxation of a nonsupersymmetric, supercritical
phase of string theory to a half-BPS state of
the critical type II theory.
In the case where the final state
is type IIA string theory, other solutions with the same
late-time asymptotics have been found \cite{sethi},
and defined nonperturbatively in the string 
coupling.\footnote{For related studies, see 
\cite{Kodama:2006bw,Ishino:2006nx,Ishino:2005ru}.}
The resolution of the early-time behavior is quite different from
ours, involving a matrix theory definition of the strong
coupling region rather than the addition of supercritical dimensions.  There are, in
other words, at least two valid completions of the
the 10-dimensional lightlike linear dilaton theory
in the strong coupling regime.  This serves as a reminder of
the dissipative nature of physics in an expanding universe.

\section{Conclusions}
We have introduced solutions to string theory that
interpolate between theories in different total numbers
of spacetime dimensions.  These backgrounds are solutions to
the classical equations of motion in the
spacetime sense, and we have shown that they are exact to all orders, 
in the worldsheet sense.  The consistency of these transitions
depends on the dynamical readjustment of the dilaton and
the metric.  These effects come from a one-loop renormalization on
the string worldsheet, and they render the total central charge on
the worldsheet critical in both limits of the solution, $X\uu +\to\pm\infty$.

The solutions work similarly in the bosonic, type 0, and 
unstable heterotic HO$^{+/}$ theories.
In the latter two cases, orbifold singularities can be introduced into the
initial state, with boundary conditions chosen such that the final state
can preserve some spacetime supersymmetry in the limit $X\uu +\to +\infty$.
Since the spacetime supersymmetry 
of the final state is an exact symmetry of the dynamics, it must
be viewed as broken spontaneously, rather than explicitly, for finite $X\uu +$.
We arrive at the striking conclusion that string theories in arbitrarily
high dimensions can preserve an exact dynamical supersymmetry, albeit
nonlinearly realized.  Our exact solutions describing transitions from
the supercritical theory to supersymmetric vacua of the
critical theory provide conclusive evidence in favor of the proposals
in \cite{hellerman1,hellerman2}.

We have shown that our CFT can be solved exactly at the classical level,
on a worldsheet of infinite extent.  It seems unlikely that quantum
corrections should vitiate the solvability of the classical theory, since
the perturbation series in all of our models terminate at one loop.  Solving the
theories explicitly in finite volume, and with quantum corrections included,
would be extremely interesting.  

The total number of string states at a given mass level goes down
when the number of spacetime dimensions decreases.
Even with our simple ansatz for classical solutions 
of the worldsheet theory, we can
see the outlines of the mechanism by which the number of
degrees of freedom in the theory is reduced.  String
states that do not `fit' into the worldsheet CFT 
of the final spacetime are pushed along the wall of
an expanding bubble.  Only string states with no
degrees of freedom excited in the massive sectors
are allowed to enter the interior of the tachyonic phase.

We also saw that the final state of the theory, including the
topology and total number of dimensions, is not dictated 
by the initial state.   Rather, it is determined
dynamically by the configuration of the closed string tachyon.  This suggests
a picture of a string landscape in which all lower-lying regions are accessible from 
a summit of highly supercritical theories.  
The notion of the supercritical string as a primordial
phase from which all other string theories can emerge can be viewed as a
closed-string extension of the
$K$-theory constructions of open string theory \cite{kwitten}.
We feel this direction of investigation deserves further development.

%
%
%


\section*{Acknowledgments}
The authors would like to thank Ofer Aharony, Juan
Maldacena, Joseph Polchinski, Savdeep Sethi, Eva Silverstein and
Edward Witten for valuable discussions.
S.H.~is the D.~E.~Shaw \& Co.,~L.~P.~Member
at the Institute for Advanced Study.
S.H.~is also supported by U.S.~Department of Energy grant DE-FG02-90ER40542. 
I.S.~is the Marvin L.~Goldberger Member
at the Institute for Advanced Study, and is supported additionally
by U.S.~National Science Foundation grant PHY-0503584. 

\bibliographystyle{utcaps}
\bibliography{dimchange}

\end{document}